\documentclass[fleqn,usenatbib]{mnras}

\usepackage{newtxtext,newtxmath}

\usepackage[T1]{fontenc}

\DeclareRobustCommand{\VAN}[3]{#2}
\let\VANthebibliography\thebibliography
\def\thebibliography{\DeclareRobustCommand{\VAN}[3]{##3}\VANthebibliography}

\usepackage{graphicx}	
\usepackage{amsmath}	
\usepackage{pdflscape} 
\usepackage{comment} 
\usepackage{bm} 
\usepackage[dvipsnames]{xcolor}
\usepackage{booktabs}

\title[Unsupervised light curve feature extraction]{Light curve fingerprints: an automated approach to the extraction of X-ray variability patterns with feature aggregation -- an example application to GRS 1915+105}
\author[J. K. Orwat-Kapola et al.]{
Jakub K. Orwat-Kapola,$^{1}$\thanks{E-mail: j.k.orwat-kapola@soton.ac.uk}
Antony J. Bird,$^{1}$
Adam B. Hill,$^{1, 2}$
Diego Altamirano$^{1}$,
\newauthor Daniela Huppenkothen$^{3}$
\\
$^{1}$School of Physics and Astronomy, University of Southampton, Southampton, Hampshire SO17 1BJ, UK\\
$^{2}$HAL24K Labs, Herikerbergweg 292, 1101 CT, Amsterdam, the Netherlands\\
$^{3}$SRON Netherlands Institute for Space Research, Sorbonnelaan 3, 3584 CA, Utrecht, the Netherlands\\
}

\date{Accepted XXX. Received YYY; in original form ZZZ}

\pubyear{2021}

\begin{document}
\label{firstpage}
\pagerange{\pageref{firstpage}--\pageref{lastpage}}
\maketitle

\begin{abstract}
Time series data mining is an important field of research in the era of ``Big Data''. Next generation astronomical surveys will generate data at unprecedented rates, creating the need for automated methods of data analysis.
We propose a method of light curve characterisation that employs a pipeline consisting of a neural network with a Long-Short Term Memory Variational Autoencoder architecture and a Gaussian mixture model. The pipeline performs extraction and aggregation of features from light curve segments into feature vectors of fixed length which we refer to as light curve ``fingerprints''. This representation can be readily used as input of down-stream machine learning algorithms.
We demonstrate the proposed method on a data set of Rossi X-ray Timing Explorer observations of the galactic black hole X-ray binary GRS 1915+105, which was chosen because of its  observed complex X-ray variability. We find that the proposed method can generate a representation that characterises the observations and reflects the presence of distinct classes of GRS 1915+105 X-ray flux variability. We find that this representation can be used to perform efficient classification of light curves. We also present how the representation can be used to quantify the similarity of different light curves, highlighting the problem of the popular classification system of GRS 1915+105 observations, which does not account for intermediate class behaviour.
\end{abstract}

\begin{keywords}
X-rays: binaries -- methods: data analysis
\end{keywords}




\section{Introduction}

\subsection{Machine learning for astronomical time series}

Automated approaches to data analysis are becoming increasingly relevant as we are entering the era of big data in and outside of astronomy. Industrial applications include, for example, analysis of smart utility meter data and city traffic data. The ability to identify appliance energy usage patterns can provide actionable insights to utility providers \citep{Singh2018}, as smart meters are being installed in millions of houses. Prediction of the rate of traffic flow and smart route planning can aid in the management of congestion in intelligent transportation systems \citep{Zhu2019}. Within the field of astronomy, future and ongoing surveys, like those conducted by Vera C. Rubin Observatory (previously known as the Large Synoptic Survey Telescope (LSST)) \citep{Ivezic2019} and Zwicky Transient Facility \citep{Bellm2014}, will produce terabytes of data at unprecedented rates, and manual analysis of this volume of data by human experts is impossible. In order to make sense of this data, analysts need methods which can extract descriptive variables from individual data observations (i.e. perform feature engineering). Resulting variables can then be used in machine learning pipelines to compare observations and perform tasks like classification, outlier detection or clustering. 

Feature engineering often requires domain specific expertise from the analyst, who needs to identify descriptors containing relevant information about the observations in the data set \citep[for example][]{Richards2011}. Automated feature extraction is an alternative to manual feature engineering, and requires less specific domain knowledge. Automated feature extraction often involves methods which encode data into a more abstract, low-dimensional, latent representation. The resulting latent representation contains compressed information about the input data, and reflects the similarities and differences between observations. Machine learning algorithms can leverage this information to perform a variety of tasks. Several methods of automated feature extraction for light curve data (time series data describing intensity of an astronomical object) have been studied in the past, for example Kohonen self-organising maps \citep{Armstrong2015} and pattern dictionary learning \citep{Pieringer2019}. \citet{Mackenzie2016} extracted light curve segments and clustered them to find common patterns of variability in variable star candidates, creating a representation compatible with machine learning classifiers. Similarly, \citet{Valenzuela2018} used a sliding window method to extract light curve segments and classified the light curves based on the presence of characteristic patterns. 
\citet{Naul2018} extracted features of phase-folded light curves using an autoencoding recurrent neural network and used a random forest classifier on observations of variable sources from All Sky Automated Survey Catalog, Lincoln Near-Earth Asteroid Research survey, and Massive Compact Halo Object Project, achieving classification accuracy of well over 90\%.\par

End-to-end architectures have also been used for a variety of tasks. Those approaches use a single model which implicitly learns a low-dimensional representation of data to perform classification, inference etc.
For example, \citet{Charnock2017} used a bi-directional recurrent neural network (RNN) to process and classify light curves of the Supernovae Photometric Classification Challenge data set and achieved impressive results. \citet{Mahabal2017} transformed the light curves of Catalina Real-Time Transient Survey to two-dimensional images and used a convolutional neural network (CNN) to classify them. \citet{Shallue2018} trained a deep CNN to detect exoplanet transits in folded light curves of the Kepler mission, and were able to discern plausible planet signals from false-positives 98.8\% of the time. \citet{Becker2020} trained an RNN to classify variable star light curves, achieving results which were competitive with results of a random forest classifier trained on features generated by a library for time series feature extraction feature analysis for time series (FATS). They grouped each light curve with a sliding window, reducing the sequence length, which allowed the RNN to process long sequences. More recently, \citet{ZhangBloom2021} demonstrated state-of-the-art classification performance using novel Cyclic-Permutation Invariant Neural Networks which are invariant to phase shifts of phase-folded light curves.\par

In this work, we contribute to the toolbox of automated characterisation methods for light curve data, as we explore the use of neural networks and Gaussian mixture models for descriptive feature generation and clustering. We construct a data representation called light curve ``fingerprints" that can be used in downstream classification, outlier detection and clustering tasks. We demonstrate that such methods can be useful in the analysis of the evolution of a particular source of interest. \par

\subsection{GRS 1915+105}

In this work, we analyse the complete data set of GRS 1915+105 observations captured by the Proportional Counter Array on-board of the Rossi X-ray Timing Explorer (RXTE/PCA) \citep{Glasser1994} between 1996 and 2011.\par

GRS 1915+105 is a galactic black hole X-ray binary system discovered in 1992 
\citep{Castro-Tirado1994}, which shows an extraordinary complexity of X-ray flux variability. It was the only known source to exhibit such intricate patterns of behaviour, until the discovery of black hole candidate IGR J17091-3624 \citep{Kuulkers2003, Capitanio2006}, which shares some of the same characteristics \citep{Altamirano2011}.\par

\begin{figure*}
	\includegraphics[width=\textwidth]{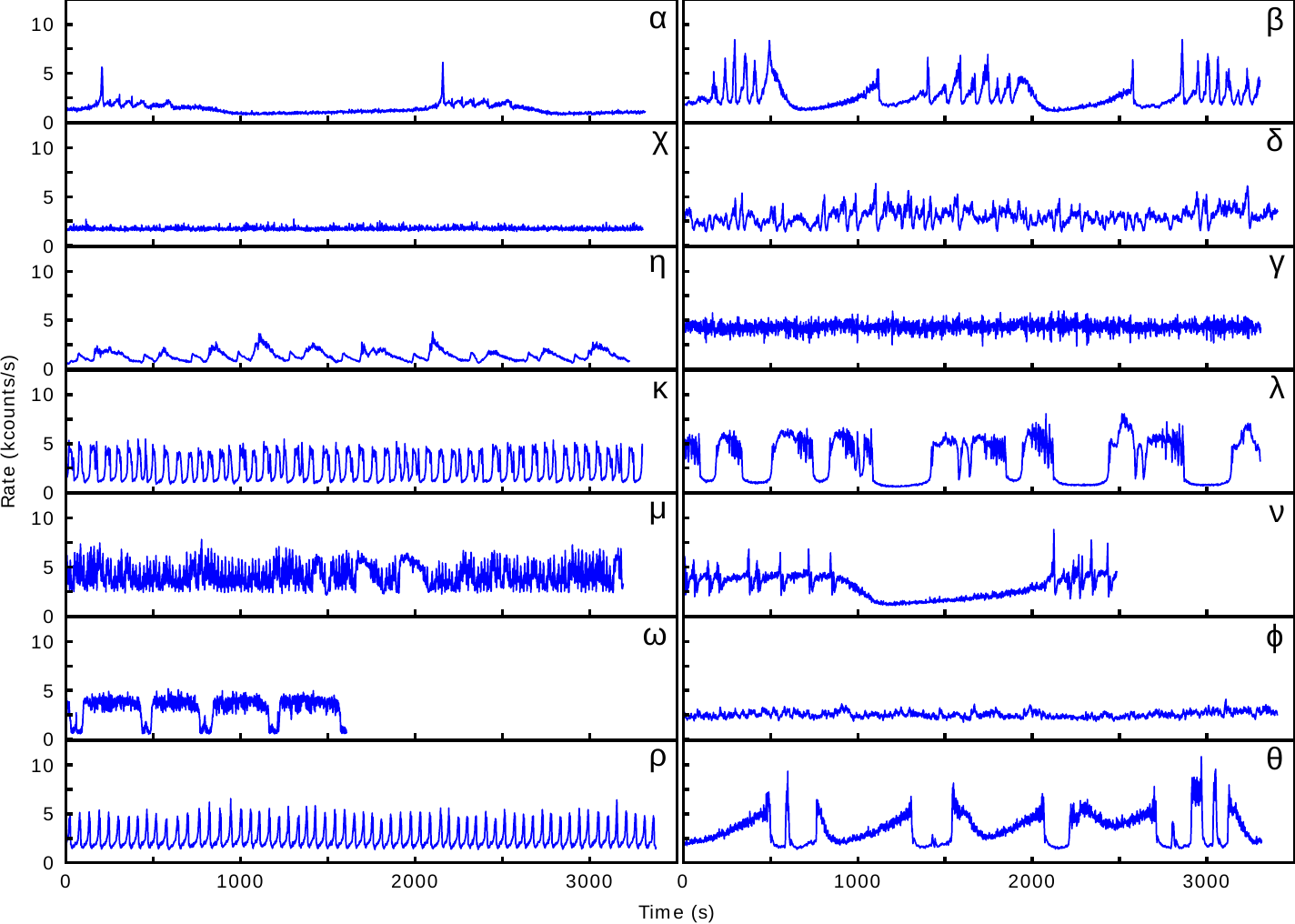}
    \caption{Examples of GRS 1915+105 X-ray light curves. Shown light curves have been classified according to the Belloni et al. system. Classifications are shown in the upper-right corner of each sub-figure. We use the curated set of classifications published by \citet{Huppenkothen2017}. Light curves have 1 second cadence.}
    \label{fig:light curves}
\end{figure*}

In an attempt to demonstrate that the complex behaviour of GRS 1915+105 is controlled by just a few simple variables, \citet{Belloni2000} constructed a classification system, which assigned observations of the source to one of 12 classes. Classification was based on the presence of qualitative patterns in light curves and colour-colour diagrams of source observations. Work that followed had expanded the classification system to the total of 14 classes \citep{Klein-Wolt2002, Hannikainen2003, Hannikainen2005}. This classification system is hereafter referred to as the ``Belloni et al. system''. Figure~\ref{fig:light curves} shows an example of GRS 1915+105 light curve for each one of the 14 classes. Some classes show steady flux without any structured variability, other show periodic flares, dips or different types of periodic and aperiodic variability. There are both inter-class and intra-class variations in the amplitude of flux variability.\par

Highly complex patterns of light curve variability of GRS 1915+105 are commonly interpreted as being caused by a partial or complete disappearance of an observable innermost region of the accretion disc. Disappearance of the disc, in turn, is caused by thermal-viscous instability of the inner region of the disc \citep{Belloni1997a, Belloni1997b}. X-ray variability patterns corresponding to the 14 classes of source behaviour can repeat almost identically months and years apart. This suggests that the instability of the disc triggers a very specific and reproducible response \citep{Belloni2001}.

GRS 1915+105 was the first discovered stellar-mass source producing highly relativistic jets \citep{Mirabel1994}. In this regard, it sparked great research interest, as it offered the possibility of studying coupled inflow-outflow processes of an accreting black hole, which unlike more massive active galactic nuclei, evolves in observable time scales \citep{Fender2004}. Plasma ejections of GRS 1915+105 in the form of jets have also been associated with the instability of the accretion disc \citep{Belloni2000, Nayakshin2000, Fender2004}, which supports the notion of disc-jet coupling.\par

Furthermore, \citet{Naik2002} found that  periods when the innermost region of the accretion disc is not observable (which are associated with variability class \(\chi\)), are preferentially followed by classes showing periodic bursts, i.e. classes \(\rho\) and \(\alpha\). Therefore, following the changes of variability classes can improve our understanding of the evolution of the source over longer time scales, and it is an important method of probing the accretion dynamics, as pointed out by \citet{Huppenkothen2017}.\par

\citet{Huppenkothen2017} conducted the first study of the whole set of GRS 1915+105 observations from RXTE/PCA using machine learning. They characterised the observations of the source and classified them according to the Belloni et al. system, using machine learning classification methods. They used features derived from power spectra, time series features extracted with an autoregressive model, and hardness ratio features derived from energy spectra.\par

\subsection{Summary of this work}

In this work, we perform classification of GRS 1915+105 observations using exclusively time series features derived from light curve data in an unsupervised manner, using a neural network. We choose not to use energy and power spectral features, found in the works of \citet{Belloni2000} and \citet{Huppenkothen2017}, hence making our method more generalisable to other data sets, sources and energy bands. In principle, any type of univariate time series data can be analysed using this method.\par

In order to perform machine learning on light curve data, they need to be represented with vectors of constant length. It is a common approach to segment light curves to create time series of constant length prior to feature extraction. However, it is not obvious how to aggregate features of a light curve from a variable number of segments. We propose the ``fingerprint'' representation as a method of aggregating segment-wise time series features into a constant-length vector which describes a whole light curve of any length, in a manner that is interpretable by machine learning algorithms. \par

We train a neural network which performs dimensionality reduction of light curve segments, producing time series feature vectors of individual segments. We then perform clustering of features of individual segments and use the number of segments assigned to each cluster to construct the ``fingerprint'' representation, taking advantage of the fact that segments showing similar patterns of light curve variability tend to cluster together. We show that the ``fingerprint'' representation can be used to quantify the similarity of light curves and to perform machine learning classification.


This paper contains the following parts: Section~\ref{sec:Data prep} describes the data preparation process, which starts with RXTE/PCA observations and produces a data set of light curve segments, a suitable input for a neural network. Section~\ref{sec:Encoding shape features} provides details about the neural network and describes the process of dimensionality reduction of the data set, which results in the encoded data representation (time series features). Section~\ref{sec: cluster analysis of segments} describes the process of cluster analysis of the encoded representation. Section~\ref{sec: 222 patterns} describes how Gaussian mixture models are used to identify the set of light curve patterns from the encoded data representation. Section~\ref{sec: fingerprints} shows how the set of light curve patterns is used to construct the observation ``fingerprints''. Section~\ref{sec: classification} shows how ``fingerprints'' can be used to assign light curves to classes of the Belloni et al. system, and Section~\ref{sec: kappa and omega} demonstrates how the ``fingerprints'' can be leveraged to refine the classification system in a data-driven way. Section~\ref{sec:conclusions} summarises the main results, discusses limitations of the presented approach and lists some ideas for further research.\par

\section{Data preparation}
\label{sec:Data prep}

We retrieve all available RXTE/PCA observations of GRS 1915+105 in Standard-1 format (0.125 second resolution light curves which combine all energy channels over the range of 2 - 60 keV) \footnote{\url{https://heasarc.gsfc.nasa.gov/db-perl/W3Browse/w3browse.pl}}. Extraction is limited to the most reliable counting array number 2 (PCU2). This yields 1776 light curves, which are subsequently re-binned. We generate two data sets: one where binning is performed at 1 second resolution, and another where binning is performed at 4 second resolution. Two data sets are generated because we fix the input size of the neural network to 128 data points due to GPU hardware constraints (see Appendix~\ref{sec: network training} for more detail), therefore the time resolution of the light curves is the main parameter which influences the amount and type of information that is provided to the network. On the one hand, short time resolution data can resolve fast variability structures of the X-ray source, but it cannot capture longer variability structures in a light curve segment of fixed size. On the other hand, longer time resolution data can capture more context within a light curve segment, but any fast variability structures are smoothed or lost. Hence, we choose to train two separate models on the two data sets and compare them in order to explore the effect of changing data resolution.\par
\begin{table}
	\centering
	\caption{Parameters of the two data sets of light curve segment.}
	\label{tab:data set parameters}
	\begin{tabular}{lcc}
		\hline
		Parameter & 4s data set & 1s data set\\
		\hline
		Cadence (s) & 4 & 1\\
		Segment length (s) & 512 & 128\\
		Stride length (s) & 8 & 10\\
		\# segments & 468,202 & 474,471\\
		\hline
		
	\end{tabular}
\end{table}

Similarly to \citet{Huppenkothen2017}, we perform light curve segmentation in order to produce a data set of segments of standard length. Only the good time intervals from each observation are segmented, and the interruption periods of missing data are skipped. We choose the segment size of 128 data points, resulting in segment length of 512 seconds for data with 4 second resolution. A moving window segmentation is performed with a stride of 2 data points (8 seconds), yielding a set of 468,202 segments derived from 1738 observations which satisfy the segmentation criteria. This data set of light curve segments is hereafter referred to as ``4s data set''. The same set of 1738 observations, binned to 1 second resolution, is segmented with a stride length of 10 data points, yielding 474,471 segments. This data set of light curve segments is hereafter referred to as ``1s data set''. Stride size is adjusted in order to make the number of segments as close as possible to the 4s data set. Table~\ref{tab:data set parameters} lists parameters of the two data sets.\par

The time duration captured by the segments is not sufficient to contain the longest cycles of flux variability produced by the source. For example, some observations of class $\alpha$ show intervals of quiescence which last $\sim$1000 seconds, and are interlaced by periods of flaring which last $\sim$500 seconds. However, the main goal of our study is not to classify individual light curve segments, but rather to construct a new, data-driven system of segment templates, which create classifiable observation signatures when grouped together with other segments from the same light curve. See Section~\ref{sec: classification} for an example  of a classification experiment which illustrates the use of observation signatures (``fingerprints'').\par

A small stride size is selected in order to maximise the number of light curve segments available for neural network training. It also ensures that the model is exposed to the full range of phase shift of light curve patterns \citep{Huppenkothen2017}.\par

Light curve segments are independently standardised; count rate values of each segment are mean centred and scaled to units of standard deviation. Segments are standardised in order to decouple their shape and intensity information. Many of the patterns observed in the variability of GRS 1915+105 repeat at various mean count rate levels. Therefore, standardisation of the segments is likely to cause segments showing similar shape patterns to align in the latent feature space extracted from the data. We also allow for the possibility that some of the shapes could be shared by several classes of variability, as described by \citet{Belloni2000}, but at different intensities, and standardisation can make it easier to draw links between those cases. \par

The resulting data set of light curve segments, together with corresponding count rate uncertainty values (needed to calculate the reconstruction error, see Equation~\ref{eq:chi-square}), is used to train a neural network. The process is described in Section~\ref{sec:Encoding shape features}.\par

Original count rate levels of the light curve segments are also important in the data analysis, so intensity information is added to the final feature set in the form of four descriptive statistics; mean, standard deviation, skewness and kurtosis, which are calculated from the distribution of count rate of each segment. These statistics make up one of the two sets of features used in cluster analysis in Section~\ref{sec: cluster analysis of segments}, and this set of four intensity features is hereafter referred to as ``intensity features of light curve segments'' (IFoS).\par

\section{Feature extraction with a neural network}
\label{sec:Encoding shape features}

In order to represent the light curve data using a small number of variables, and hence allow us to analyse that data space in a relatively small number of dimensions, we perform dimensionality reduction using a neural network. This process aims to extract a small set of ``hidden'' variables, which encode information about the structured variability of the light curves. Various methods of modelling of time series data have been employed in numerous fields of research \citep{Langkvist2014, Hyndman2016, Benkabou2018, IsmailFawaz2019}.\par

In order to address the problem of dimensionality reduction of light curve data, we propose a Variational Autoencoder (VAE) with Long-Short Term Memory (LSTM) cells within its encoder and decoder blocks. VAE is a type of probabilistic neural network model first proposed by \citet{Kingma2014}. As mentioned, it uses two blocks of neurons; an encoder, which maps input data onto a small set of distributions (commonly referred to as continuous latent variables), and a decoder, which samples from those distributions and maps them back to the input data space, hence reconstructing observations of input data. VAE is therefore trained to effectively perform compression and decompression of input data. The compression process requires the construction of a meaningful latent space of considerably smaller dimensionality than the input. Resulting latent variables are the compressed representation of data and can be leveraged in the data analysis process. \par

Autoencoders have been used in the past for the purpose of feature extraction from astronomical data \citep[for example][]{Naul2018}. Here we use a variational variant of an autoencoder, because it introduces a Gaussian prior which applies a regularization effect on the produced latent space. Non-regularized autoencoders can map similar inputs to very different latent variables \citep{HintonLecture}, therefore regularization of the latent space is required for the down-stream clustering and merging of clusters discussed in Section~\ref{sec: cluster analysis of segments}.

LSTM cells found in the encoder and decoder blocks of the proposed architecture are a type of recurrent neural network, first proposed by \citet{Hochreiter1997}. RNNs learn from sequential data, and their use has been researched extensively for the processing of text, handwriting, speech and sound \citep[and references therein]{Yu2019}.\par

Cells of an RNN process sequential input one data point at a time. At every iteration, the state of the cell is fed as input of the next iteration, which allows the network to learn from consecutive points of the sequence and extract temporal patterns. RNNs are able to make predictions based on the immediate context of the processed data point, but they tend to quickly forget information that is not frequently reinforced, which means that they struggle to capture long term patterns. LSTM cells address this issue through the introduction of so called ``gates'' which consist of non-linear functions that control the state of the LSTM cell and protect the relevant information over long time scales.\par

\subsection{Training of the neural network}
\label{sec: training network}

Both data sets are subdivided into training, validation and testing subsets. In order to ensure that the subsets are independent, segments derived from the same observation are only included in one of the subsets. In order to ensure completeness of the subsets, observations which have Belloni et al. system classifications available \citep{Huppenkothen2017}, are assigned to the subsets in a random, stratified fashion. At least one observation of each class is assigned to each subset, and the remaining observations are assigned to training, validation and testing subsets according to the split ratio of 7/1/2. Since only two observations of \(\eta\) class are available at this stage, both observations are assigned to the training subset. Observations without Belloni et al. system classifications are randomly distributed between the three subsets, while accounting for the fact that observations contain variable number of count rate data points. The resulting training sets contain $\sim$70\% of total data points, validation sets contain $\sim$10\% of data points, and testing sets contain $\sim$20\% of data points.

Two LSTM-VAE neural networks with identical architecture are trained to compress light curve segments of 128 data points into 20 continuous latent variables, one network for each data set. Data from the training set is used to adjust the parameters of the networks, and the validation set is used to measure whether the adjustment improved the networks' ability to process previously unseen examples of data. Testing set is kept aside until all training and fine-tuning of models is finished. We perform dimensionality reduction on both data sets of standardised light curve segments, which are generated using the method described in Section~\ref{sec:Data prep}).\par

The networks are built using Keras, an open-source neural network library \citep{chollet2015keras} with Tensorflow backend, and Python 3 programming language\footnote{relevant code for data pre-processing and model training will be available at \url{https://github.com/jorwatkapola/autoencoders-GRS-1915}}. More details about network architecture can be found in Appendix~\ref{sec: network architecture}.

Output of the network consists of the reconstruction of an input light curve segment. Performance of the network is quantified using a loss function, which depends on the reconstruction error and a regularisation term. The loss function is a sum of a mean square error weighted by the data uncertainty (Equation~\ref{eq:chi-square}) and Kullback-Leibler divergence (Equation~\ref{eq:KL-divergence}). The weighted mean square error is defined as:
\begin{equation}
    \text{WMSE} = \frac{1}{N}\sum_{i=1}^N \left( \frac{x_i-\Hat{x_i}}{\sigma_i} \right)^2
	\label{eq:chi-square}
\end{equation}
where \(N\) is the number of processed data points, \(x_i\) is the value of a data point \(i\) in the input light curve segment, \(\Hat{x_i}\) is the value of data point \(i\) in the reconstructed sequence, and \(\sigma_i\) is the uncertainty of the value of \(i\) in the input light curve sequence. Uncertainty values are scaled with the same value of standard deviation as the count rate values of light curve segments.\par

Kullback-Leibler divergence is a regularisation term, which prevents the distribution of the latent variables from substantially deviating from the standard normal distribution. This term is used to ensure that the latent space is continuous and meaningful. Kullback-Leibler divergence is defined as:

\begin{equation}
    D_{\text{KL}} = -\frac{1}{2Z}\sum_{j=1}^Z \left( \log(\Hat{\sigma_j}^2) - \Hat{\mu_j}^2 - \Hat{\sigma_j}^2 +1 \right)
	\label{eq:KL-divergence}
\end{equation}
where \(Z\) is the number of latent variables, \(\Hat{\sigma_j}^2\) is the variance of latent variable \(j\), and \(\Hat{\mu_j}\) is the mean of latent variable \(j\).\par

\begin{figure*}
	\includegraphics[width=\textwidth]{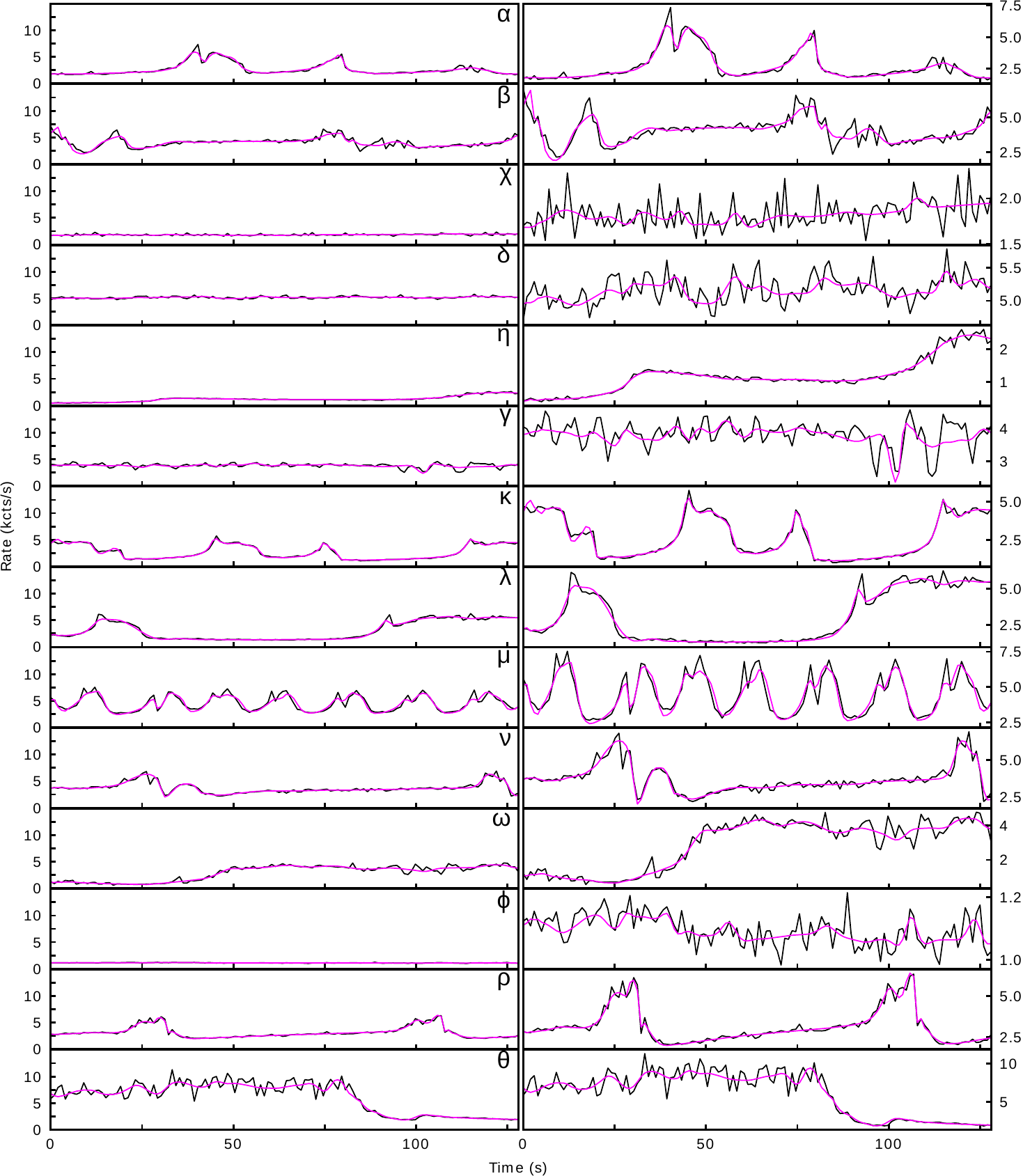}
    \caption{Examples of light curve segments from 1s data set. Segment reconstruction output of the LSTM-VAE is shown in magenta. Segments originate from observations that had been classified according to the Belloni et al. system. We use the curated set of classifications published by \citet{Huppenkothen2017}. Both columns shows the same segments, but in the right column the range of the vertical axis is dynamic. All segments come from the testing subset of 1s data set, with exception of class \(\eta\), which was included only in the training subset. 
    }
    \label{fig:segment reconstructions 1s}
\end{figure*}

The networks are trained to minimise the value of the loss function. Details about the training process of the neural networks can be found in Appendix~\ref{sec: network training}. Examples of light curve segments from 1s data set, together with their LSTM-VAE reconstructions are shown in Figure~\ref{fig:segment reconstructions 1s}. Examples for segments from 4s data set are shown in Appendix~\ref{sec:segment reconstructions 4s}. \par


\subsection{Light curve feature extraction}

The network infers mean and variance of the 20 continuous latent variables from the data  (i.e. outputs of \(\texttt{Latent\_mean}\) and \(\texttt{Latent\_log\_variance}\) layers of the network correspond to the mean and variance parameters, see Appendix~\ref{sec: network architecture} for more details). The latent variables are a compressed representation of the network's input. For the purpose of further analysis, we do not sample from latent distributions, but take the means of latent distributions, which are representative of the position of each light curve segment within the latent space. The resulting set of latent variables is hereafter referred to as ``shape features of light curve segments'' (SFoS). Hence, the shape information of each light curve segment is represented by 20 SFoS values.\par

In order to assess how well SFoS describe the shape of light curve segments, we perform reconstruction of several segments. Figures~\ref{fig:segment reconstructions 1s} and \ref{fig:segment reconstructions 4s} show reconstructions of light curve segments of each class of the Belloni et al. system. This set of segments demonstrates how the LSTM-VAE responds to a range of different light curve patterns found in the data sets. The model is able to reproduce the gross features of each segment, but it often does not account for fast count rate changes, which results in reconstructions which are significantly smoother than the input segments. This means that SFoS likely do not account for differences between segments lacking structured variability, where the major difference lies in the root mean square (RMS) deviation from the mean. Reconstructions of those segments would differ only in terms of the shape of random noise. For example, see segments of class \(\phi\) and \(\chi\) in Figure~\ref{fig:segment reconstructions 4s}.\par

In order to remedy this limitation, IFoS (containing information about count rate mean, standard deviation, skewness and kurtosis of each segment) are used in the cluster analysis stage (Section~\ref{sec: cluster analysis of segments}) together with SFoS. Segments with indistinguishable shapes and dissimilar RMS can be distinguished based on their standard deviation values.\par

\begin{figure}
	\includegraphics[width=\columnwidth]{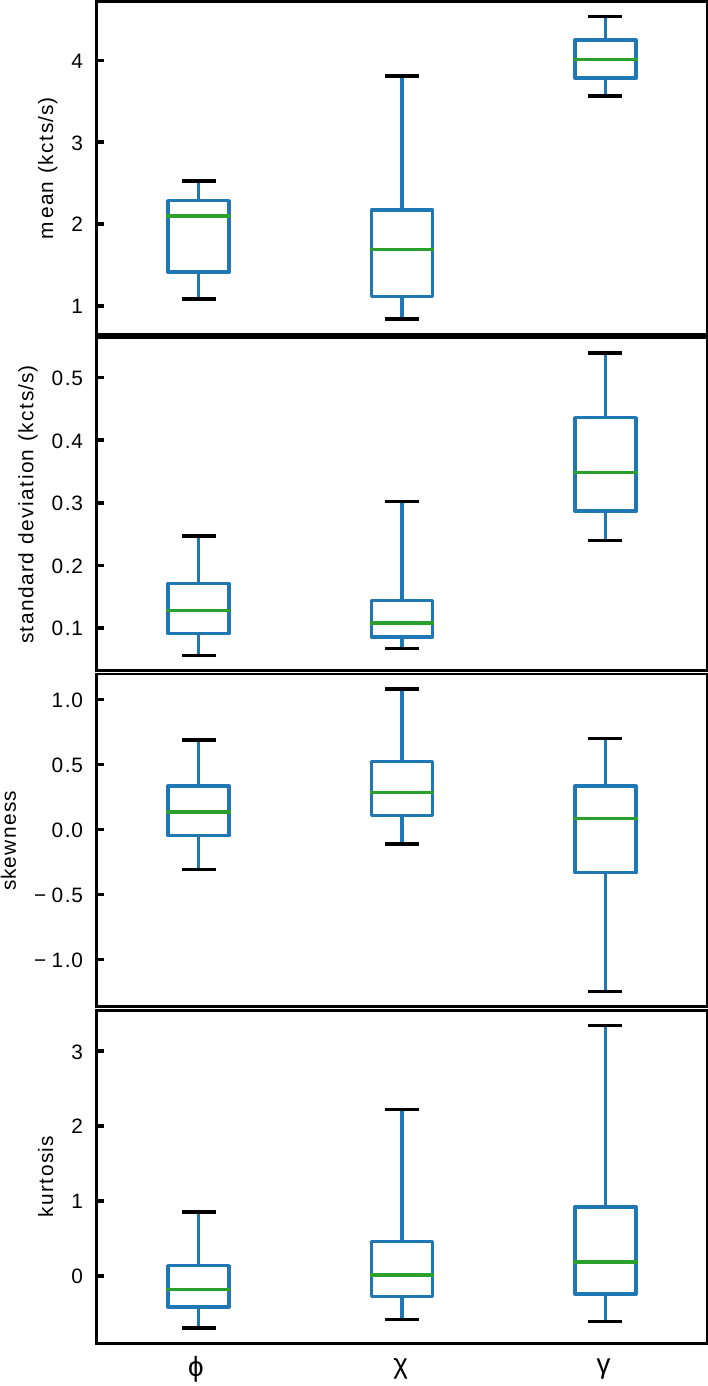}
    \caption{Distributions of the IFoS values from segments assigned to classes \(\phi\), \(\chi\) and \(\gamma\) of the Belloni et al. system. Classifications come from the set curated by \citet{Huppenkothen2017}. Four IFoS are the mean, standard deviation, skewness and kurtosis of count rate values of each segment. Box extends from the first to third quartile, and the green line shows the median. Whiskers extend from the box to the 5th and 95th percentiles.}
    \label{fig:stat boxplot}
\end{figure}

\begin{figure}
	\includegraphics[width=\columnwidth]{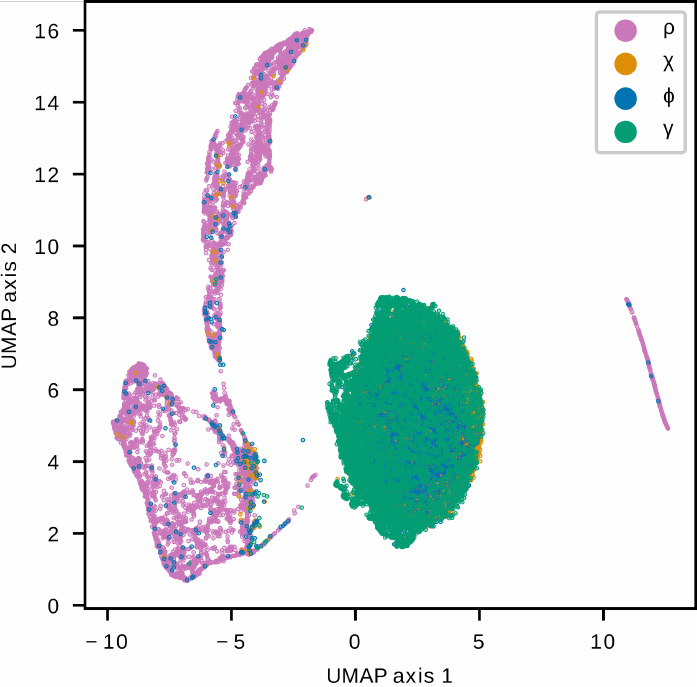}
	\includegraphics[width=\columnwidth]{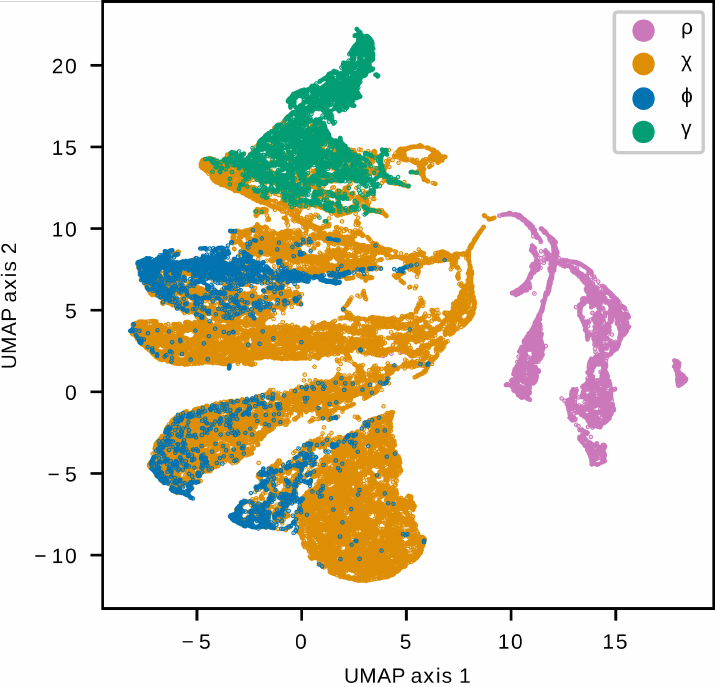}
    \caption{UMAP projection of SFoS (top) and IFoS (bottom) of light curve segments from the 1s data set which have Belloni at al. system classifications of \(\phi\), \(\chi\), \(\gamma\) or \(\rho\). Each point represents a light curve segment, and colour-coding shows their classification according to the Belloni et al. system.}
    \label{fig:SFoS and IFoS UMAP}
\end{figure}

Figure~\ref{fig:stat boxplot} shows the distribution of IFoS values for segments of \(\phi\), \(\chi\) and \(\gamma\) classes from the 1s data set. As expected, segments of class \(\gamma\) generally have larger mean and standard deviation values than the other two classes, which indicates that IFoS would allow for segments of class \(\gamma\) to be distinguished from classes \(\phi\) and \(\chi\), even in segments where the characteristic ``dip'' of class \(\gamma\) is not observed.\par

Furthermore, projections of SFoS and IFoS in Figure~\ref{fig:SFoS and IFoS UMAP} show that IFoS could be used to distinguish classes \(\phi\), \(\chi\), and \(\gamma\) much more reliably than SFoS alone. The projection was produced using Uniform Manifold Approximation and Projection (UMAP) \citep{McInnes2018} algorithm, which aims to preserve the global structure of the high-dimensional data. In the projection of SFoS, data classified as \(\phi\), \(\chi\), and \(\gamma\) occupied the same region of UMAP space (in the central cluster of the top sub-figure), indicating that those classes are mostly indistinguishable in the SFoS space. Data classified as \(\rho\) tends to occupy separate regions of the SFoS space. Class \(\rho\) is included in the projection to demonstrate that data associated with characteristic light curve shapes tends to be distinguishable in terms of its position in the SFoS space. IFoS projection uses the same subset of light curve segment data. Classes still show significant overlap in the IFoS projection, but intensity features can clearly provide meaningful information about the differences between classes \(\phi\), \(\chi\), and \(\gamma\).\par

\section{Cluster analysis of generated features}
\label{sec: cluster analysis of segments}

\subsection{Identifying the set of light curve patterns}
\label{sec: 222 patterns}

Sections~\ref{sec:Data prep} and \ref{sec:Encoding shape features} describe the process of feature engineering of four IFoS and 20 SFoS, which encode the shape and intensity information about light curve segments. This set of 24 features is hereafter collectively referred to as ``shape and intensity features of light curve segments'' (SIFoS). We generate two separate sets of SIFoS, one for the 1s data set, and one for the 4s data set.\par

In order to find the exhaustive set of light curve pattern templates which have been produced by GRS 1915+105, we perform clustering of the data in the 24 dimensional space of SIFoS. Clustering is performed with an implementation of Gaussian mixture model (GMM) included in the machine learning library scikit-learn \citep{Pedragosa2011}. The algorithm approximates the probability distribution of the data using a set of multidimensional Gaussian components (see Appendix~\ref{sec: gaussian mixture model} for more details).\par

We explored the possibility of using density-based clustering algorithms DBSCAN \citep{Ester1996} and OPTICS \citep{Ankerst1999}, and we find that it is difficult to fine-tune the hyper-parameters of those algorithms in a way that would not lead to a large proportion of the data being rejected as noise. We find that GMM offers a much more straightforward method of assigning the data to clusters. Therefore, we focus on the discussion of the application of GMM in the remainder of this text.

We choose to use GMM to approximate the shape of latent manifold in the SIFoS space, with the intention of merging of the Gaussian components which show significant overlap. This way we use combinations of Gaussian components to account for the presence of any extended, curved, non-Gaussian structures in the latent space. We are assuming that those structures correspond to particular light curve patterns, and for each observation of the source, the relative amount of time the source spends showing those patterns allows us to determine the class of that observation.

\begin{figure}
	\includegraphics[width=\columnwidth]{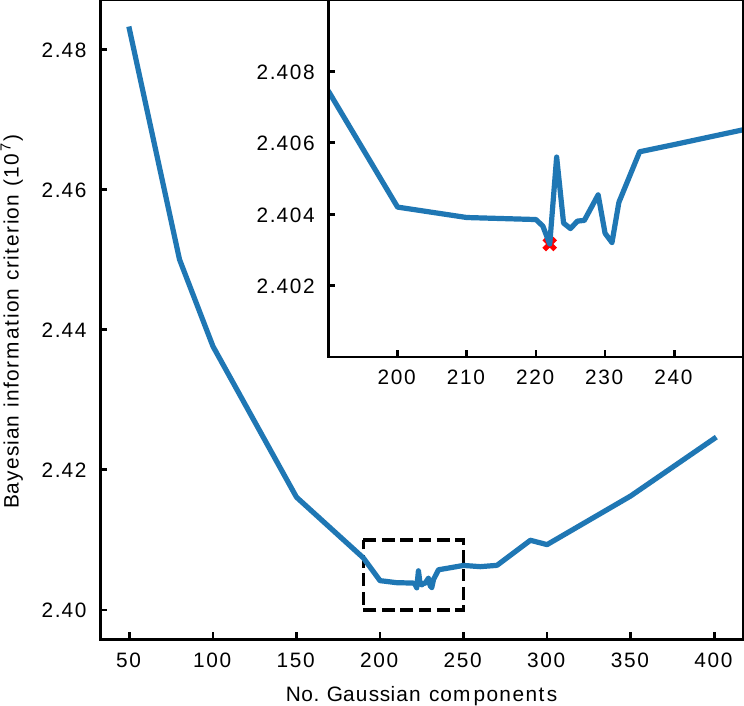}
	\includegraphics[width=\columnwidth]{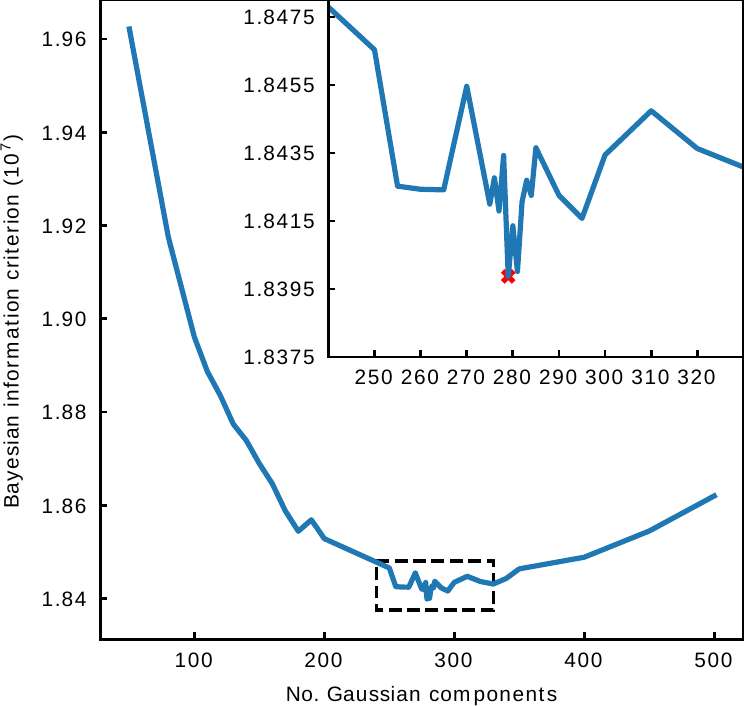}
    \caption{Bayesian information criterion of GMM as a function of the number of Gaussian components. We performed grid searches for the 1s data set (top) and the 4s data set (bottom). Figure insets are zoomed into the regions indicated by the black dotted line boxes, which contain the global minima. Minima were found at 222 Gaussian components for 1s data set and at 279 Gaussian components for 4s data set. Minima are marked by red crosses.}
    \label{fig:BIC grid search}
\end{figure}

We perform a hyper-parameter grid search to find the optimal number of Gaussian components for the GMMs in SIFoS space. Figure~\ref{fig:BIC grid search} shows values of the Bayesian information criterion (BIC) of those models as a function of the number of Gaussian components. BIC is a criterion commonly used to select the best model from a set of models fit to the same data set. The number of free parameters is one of the terms of BIC, so it penalises overly complex models. Model which produces the minimum BIC achieves the compromise between complexity and likelihood of the data (see Equation~\ref{eq:BIC}).

\begin{equation}
    \text{BIC} = k \cdot \log(N) -2 \cdot \mathcal{L}
	\label{eq:BIC}
\end{equation}

where $k$ is the number of free parameters of the model, $N$ is the number of samples of data, and \(\mathcal{L}\) is the log-likelihood of the data.\par

Grid searches indicate that the global BIC minimum for 1s data set is produced by a models with 222 Gaussian components, and for 4s data set with 279 components. We accept those numbers as the optimal numbers of components for GMM, however stochastic nature of the algorithm could cause the numbers to change slightly if the grid-search was to be repeated. The set of clusters resulting from the assignment of light curve segment data points to one of the Gaussian components of GMM is hereafter referred to as ``Gaussian clusters''.\par

Light curve segments showing similar type of variability patterns and count rate distribution are expected to produce similar values of SFoS and IFoS. Consequently, segments showing similar patterns are separated by smaller distances within the SIFoS feature space. Therefore, it is expected that Gaussian clusters contain homogeneous subsets of light curve segments, which are more similar to each other than to segments found in other Gaussian clusters.\par

Qualitative inspection of populations of the 222 Gaussian clusters of the 1s data set revealed that it is the case indeed. 42 Gaussian clusters contain recurrent flares, characteristic of light curve classes \(\rho\) and \(\nu\) (see Figure \ref{fig:segment reconstructions 1s} for examples). Table~\ref{tab:classes in clusters} shows a breakdown of the population of the clusters in terms of the classes of Belloni et al. system, and it shows that the Gaussian clusters containing recurrent flares are indeed dominated by classes \(\rho\) and \(\nu\). 69 Gaussian clusters contain segments with low RMS and no obvious patterns of variability, and their population consist mostly of classes \(\chi\) and \(\phi\). Further 15 Gaussian clusters contain segments showing similar type of behaviour, but at higher average count rates, and the population of those clusters consist mostly of classes \(\chi\), \(\gamma\) and \(\theta\). Virtually every one of the remaining Gaussian clusters has some characteristic pattern; irregular flares, dips, negative or positive gradient, flaring followed by quiescence and vice versa. Few Gaussian clusters contain segments whose common patterns of variability are not immediately apparent upon visual inspection of a small random sample of segments, which can indicate that the number of Gaussian components of the GMM is too small.\par

In order to find the minimal, exhaustive set of light curve variability patterns, segments showing the same characteristic patterns should arguably all belong to one cluster. Gaussian clusters produced with the 222 and 279 component GMMs contain apparent degeneracies. The presence of very similar Gaussian clusters is caused by the limitation of the GMM, which approximates the probability density of the data set using multivariate Gaussian components. A single component cannot spread over a curved data manifold; only multiple components can approximate the curvature, as a series of locally flat sections. Light curve segments which show a similar type of variability pattern can vary in more than one SIFoS due to their non-linear interaction within the neural network model. Therefore, segments which, for example, vary in the frequency and amplitude of a similar type of pattern, can follow a curved manifold, and hence end up in separate Gaussian clusters. We address that in Section~\ref{sec: classification}.\par

\begin{figure}
	\includegraphics[width=\columnwidth]{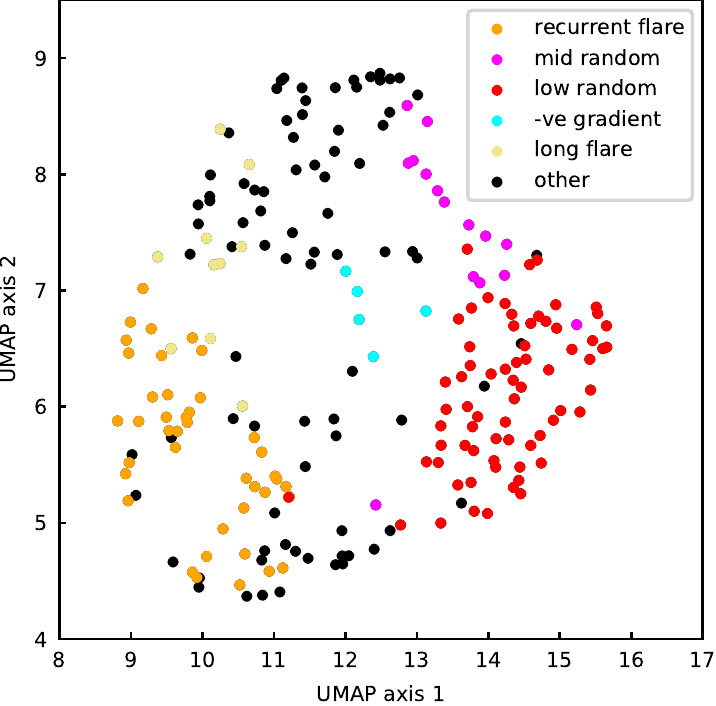}
    \caption{UMAP projection of the 222 means of Gaussian components of the GMM fit to 1s data set. UMAP reduced dimensionality from 24 to 2. Some points are colour coded, and common colours indicate Gaussian clusters containing light curve segments which show common characteristic variability patterns. As mentioned in the text, ``recurrent flare'' stands for behaviour characteristic of light curve classes \(\rho\) and \(\nu\), ``mid random'' and ``low random'' stands for the lack of apparent structured variability (only random noise) at low and medium-high mean count rates, ``-ve gradient'' stands for slow decrease of mean count rate, whereas ``long flare'' stands for irregular, long periods of flaring characteristic of classes \(\kappa\) and \(\lambda\). All of the remaining Gaussian clusters are labelled as ``other''. The population of each type of cluster in terms of the classes of Belloni et al. system is shown in Table~\ref{tab:classes in clusters}.}
    \label{fig:UMAP component means}
\end{figure}
 \begin{table*}
  \caption{Percentage distribution of light curve segments across the classes of Belloni et al. system for each type of Gaussian clusters identified in Figure~\ref{fig:UMAP component means}. Classes showing similar patterns tend to be clustered together, which is an indication of the meaningfulness of the SIFoS feature space.}
  \label{tab:classes in clusters}
\begin{tabular}{lrrrrrrrrrrrrrr}
\hline
 &  \(\alpha\) &  \(\beta\) &  \(\chi\) &  \(\delta\) &  \(\eta\) &  \(\gamma\) &  \(\kappa\) &   \(\lambda\) &   \(\mu\) &   \(\nu\) &   \(\omega\) &   \(\phi\) &   \(\rho\) &  \(\theta\) \\
\hline
recurrent flare &    1.2 &   1.0 &   0.0 &    0.0 &   0.0 &    0.0 &    0.9 &     0.0 &  0.1 &  5.9 &    0.0 &   0.0 &  90.9 &    0.0 \\
\midrule[0.05pt]
mid random      &    0.1 &   2.4 &  46.9 &    0.8 &   0.0 &   27.5 &    0.3 &     0.1 &  0.0 &  0.9 &    3.9 &   1.0 &   0.0 &   16.1 \\
\midrule[0.05pt]
low random      &    2.5 &   2.5 &  80.4 &    0.0 &   1.4 &    0.0 &    0.0 &     0.4 &  0.0 &  1.3 &    0.0 &  11.0 &   0.0 &    0.4 \\
\midrule[0.05pt]

-ve gradient    &   15.5 &  19.9 &   0.1 &   12.2 &  14.4 &    0.1 &    0.0 &     3.4 &  0.2 &  2.8 &    0.1 &   4.8 &   0.0 &   26.7 \\
\midrule[0.05pt]

long flare      &    0.5 &   9.7 &   0.0 &    2.8 &   0.0 &    0.0 &   82.3 &     1.1 &  1.1 &  0.0 &    0.0 &   0.1 &   2.3 &    0.1 \\
\midrule[0.05pt]

other           &    4.1 &  11.3 &  22.6 &    8.6 &   2.4 &    8.2 &    7.3 &     4.0 &  7.4 &  3.5 &    0.9 &   1.9 &   3.4 &   14.5 \\
\hline
\end{tabular}
 \end{table*}

Figure~\ref{fig:UMAP component means} shows a two-dimensional UMAP projection of mean positions of GMM Gaussian components which are fit to the 1s data set. Some of the points are colour coded to indicate the characteristic patterns of light curve segments found in the corresponding Gaussian clusters. Colour coding is produced by manual inspection of data found in Gaussian clusters, which is not a part of the proposed method of unsupervised data analysis. The purpose of this visualisation is to shows that Gaussian clusters sharing the characteristic patterns of behaviour tend to occupy similar regions of the SIFoS space.\par

\subsection{Relating the set of light curve patterns to the Belloni et al. system}
\label{sec: fingerprints}

In order to transform the two sets of clusters into a sets of observation features, for each observation we count the number of light curve segments assigned to each of the Gaussian components. This results in 222 values per observation in 1s data set and 279 values per observation in 4s data set. For each observation we divide the counts by the sum of all counts for that observation. Feature vectors are independently normalised in order to reduce the impact of variance in the total number of segments extracted from each observation. This variance is caused by the fact that observations vary in total duration and the number of data gaps. Since feature vectors are normalised, they only contain information about the relative abundance of light curve patterns within the corresponding observation. We refer to such 222-vectors and 279-vectors as observation ``fingerprints'', because they allow identification of distinct classes of light curve variability. \par

\begin{figure*}
	\includegraphics[width=\textwidth]{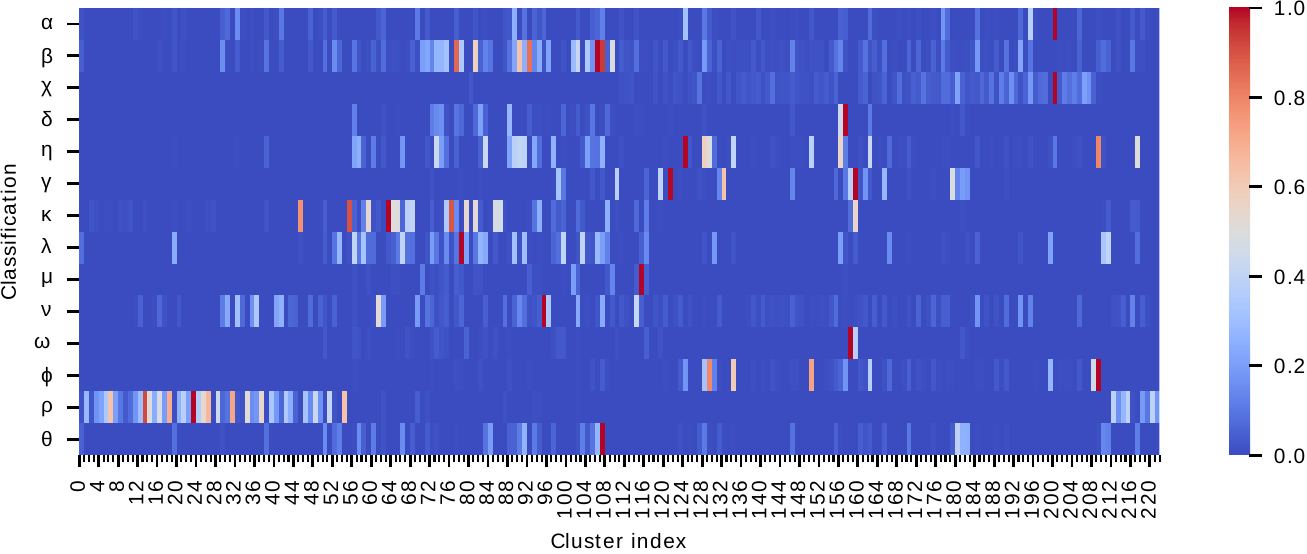}
    \caption{Heat map showing the distribution of light curve segments across the set of 222 Gaussian clusters which were fit to the classified subset of 1s data set. Heat map was normalised row-wise. Colour indicates the relative abundance of light curve segments in the corresponding cluster. Clusters are ordered based on their proximity in the SIFoS space; this was determined using a hierarchical (single linkage) clustering algorithm. Observations' class labels come from \citet{Huppenkothen2017}.}
    \label{fig:GMM vs Belloni heatmap}
\end{figure*}

In order to showcase the usefulness of ``fingerprint'' representation of data, Figure~\ref{fig:GMM vs Belloni heatmap} shows the subset of 1s data set which has been human-labelled according to the Belloni et al. system, in terms of the Gaussian clusters described in Section~\ref{sec: cluster analysis of segments}. Figure~\ref{fig:GMM vs Belloni heatmap} shows combined ``fingerprints'' for each class of observation (``fingerprints'' of observations of the same class were summed to produce the combined ``fingerprints''). Rows of this heat map correspond to the 14 classes of observations, and columns correspond to the 222 Gaussian clusters of light curve segments. Particular cells of the heat map reflect the relative abundance of segments of a particular class which have been assigned to a particular Gaussian cluster. Heat map was normalised row-wise, which means that high values indicate the Gaussian clusters which are most closely associated with a particular light curve class. This in turn indicates which light curve patterns are most abundant in light curves of a particular class.\par

The distinct appearance of the rows of the heat map indicates that the representation of light curves in terms of Gaussian clusters allows us to distinguish observations of different classes of the Belloni et al. system. This indicates that the light curve representation which employs the set of patterns could serve as a viable feature space for supervised classification algorithms.\par

\subsection{Classification of light curves using ``fingerprint'' representation}
\label{sec: classification}

In order to test the usefulness of the Gaussian cluster representation in light curve classification, we train random forest classifiers \citep{Breiman2001} to assign observations to the Belloni et al. system based on their ``fingerprints'' (note: we choose to disregard the subdivision of the \(\chi\) class, because the main distinguishing feature of those sub-classes is the position in the colour-colour diagram). We perform classification with the random forest classifier implementation included in the machine learning library scikit-learn \citep{Pedragosa2011} (see Appendix~\ref{sec: random forest classifier} for more details about the algorithm). We train separate classifiers for 1s data set and 4s data set. \par

In order to address the issue of degeneracy of light curve patterns described in Section~\ref{sec: 222 patterns} we merge Gaussian clusters separated by small distances. We use the Mahalanobis distance metric between the mean positions of Gaussian components. Mahalanobis distance is a distance metric used to measure the distance between a point and a distribution, while scaling the distance using the variance of that distribution. Mahalanobis distance is defined as:
\begin{equation}
    D_{\text{M}} = \sqrt{(\bm{u}-\bm{v})\bm{\mathsf{V}}^{-1}(\bm{u}-\bm{v})^{T}}
	\label{eq:Mahalanobis distance}
\end{equation}
where $u$ and $v$ are vectors whose separation is calculated, and $V$ is the co-variance matrix.\par

The distance threshold between means of Gaussian components which are merged is one of the hyper-parameters included in the grid-search during the training of random forest classifiers. Gaussian components are allowed to merge if the distance calculated for both co-variance matrices is smaller than the threshold hyper-parameter.\par

\begin{table}
	\centering
	\caption{List of hyper-parameters included in the grid-search of classification experiment described in Section~\ref{sec: classification}. Hyper-parameters criterion and max\_depth belong to the random forest classifier. The possible values of criterion are “gini” and “entropy”, which stand for Gini impurity and information gain respectively (see Appendix~\ref{sec: random forest classifier} for more details), and max\_depth controls the depth of decision trees of the random forest. ``Merge distance'' refers to the Mahalanobis distance threshold used in the process of merging Gaussian clusters. We test 100 values evenly spaced between 1.5 and 5. Hyper-parameter producing the largest F1 values for the two data sets are also listed.}
	\label{tab:random forest hyperparameters}
	\begin{tabular}{lccc}
		\hline
		Hyperparameter & Possible values & 1s data set & 4s data set\\
		\hline
		criterion & ``gini'', ``entropy''  & ``gini'' & ``gini''\\
		max\_depth & None, 5, 10, 15, 25  & 5 & 15\\
		merge distance & between 1.5 and 5 & 3.34 & 2.84\\
		\hline
	\end{tabular}
\end{table}

We use the training and validation data subsets (described in Section~\ref{sec: training network}) for the training of random forest classifiers. We find that the label of observation 10258-01-10-00 from \citet{Huppenkothen2017} disagrees with preceding literature \citep{Klein-Wolt2002, Belloni2013}, therefore we change the label from \(\lambda\) to \(\mu\) prior to classifier training.\par

Due to the limited number of labelled observations, we do not perform n-fold cross-validation, but instead we train the random forest classifier on 137 observations randomly sampled in a stratified manner from the data set of 159 observations which were used for training and validation of the neural network, and use the remaining 22 for validation. We repeat this process 100 times for each combination of hyper-parameters and find the mean of performance scores resulting from the 100 validation trials. In order to account for the class imbalance, the classifiers automatically adjust weights of each training sample to be inversely proportional to class frequencies. Table~\ref{tab:random forest hyperparameters} lists hyper-parameters included in the grid-search and Appendix~\ref{sec: random forest classifier} contains more details about the training of random forest classifier.\par 

We use F1 and accuracy scores to measure the performance of classifiers. Both scores can take values in the range between 0 and 1, the higher the better. Accuracy is the proportion of correct classifications out of all predicted classifications. F1 score is the harmonic mean of recall and precision of classifications of a single class:

\begin{equation}
    F_1 = 2 \cdot \frac{\text{precision} \cdot \text{recall}}{\text{precision}+\text{recall}}
	\label{eq:F1}
\end{equation}
where recall is the proportion of true positives out of the sum of all positives, and precision is the proportion of true positives out of the sum of true positives and false negatives.\par

Reported F1 scores are weighted averages across the 14 classes. The scores are weighted by the number of observations of the corresponding class in order to account for the class imbalance. In general, weighted F1 is a more reliable performance indicator, because accuracy can be easily biased when observations of one class significantly outnumber observations of other classes.\par

We find that the highest average validation performance scores for the 1s data set are weighted F1 of \(0.814\pm0.065\) and accuracy of \(0.854\pm0.054\), while the highest average validation scores for the 4s data set are \(0.760\pm0.068\) and \(0.810\pm0.056\) (reported uncertainty values are equal to one standard deviation calculated from the performance scores of the 100 validation trials). Therefore, we conclude that features derived from 1s data set perform better in the task of light curve classification. Hyper-parameters producing the highest average validation scores are listed in Table~\ref{tab:random forest hyperparameters} for both data sets.\par

A random forest classifier with the optimal set of hyper-parameters is trained on the observations of the training subset and tested on the testing subset of 1s data set, containing 47 observations. Random initiation of the algorithm is a significant cause of variance in the model performance, therefore training and testing is repeated one thousand times. Mean of weighted F1 and accuracy performance scores of those classifications are \(0.878\pm0.027\) and \(0.894\pm0.027\) respectively. It should be noted that reported uncertainty values account only for variance caused by changes to the initial random state of the classifier. \par 


\begin{figure}
	\includegraphics[width=\columnwidth]{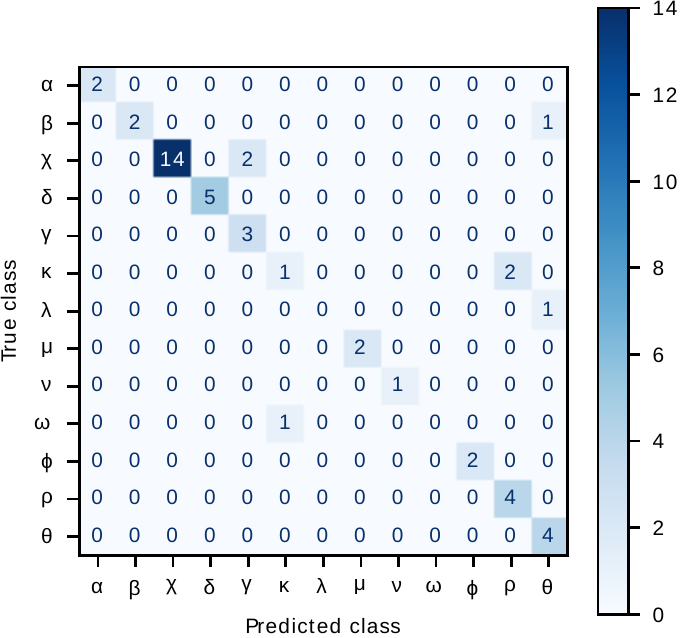}
    \caption{Confusion matrix showing classification results produced by the random forest classifier with optimal set of hyper-parameters for 47 testing set observations. The matrix shows results for the lowest weighted F1 score out of the one thousand random initiations of the classifier. Weighted F1 and accuracy of this initiation are 0.834 and 0.851 respectively. The mean weighted F1 and accuracy performance scores across the one thousand initiations are \(0.878\pm0.027\) and \(0.894\pm0.027\) respectively}
    \label{fig:confusion matrix}
\end{figure}

Figure~\ref{fig:confusion matrix} shows the classification results with the lowest performance scores out of the set of 1,000 testing trials. We present the results of the test which achieved the lowest scores, because they reveal the most information about the low-confidence classifications. It appears that our classifications disagree the most with the human-assigned labels for classes that have few observations available. Therefore it is likely that classification performance could improve given a larger amount of training data. Since the number of labelled observations is very small, the variance of test results is likely to be high. In order to test this, for each combination of hyper-parameters shown in Table~\ref{tab:random forest hyperparameters}, we perform 100 tests where 69 test observations are randomly sampled in a stratified manner from the entire data set of 206 labelled observations (combined training, validation and test data subsets), and the remaining observations are used for training. Appendix~\ref{sec: validation classifications} shows the aggregated classification results of the 100 tests of the classifier with optimal hyper-parameters. Those results seem to agree with the results shown in Figure~\ref{fig:confusion matrix}. The mean of weighted F1 and accuracy performance scores of those classifications are \(0.877\pm0.031\) and \(0.896\pm0.026\) respectively. The results also show that classes which have only one training observation available (\(\lambda\), \(\nu\) and \(\omega\)) tend to disagree with the human-assigned labels the most frequently. \par

From the test results presented in Figure~\ref{fig:confusion matrix}, we examine each observation whose classification disagrees with the human-assigned labels. Observations of class \(\beta\) get assigned to other classes most likely because of the complex behaviour of its light curves. Some of the patterns of class \(\beta\) can be seen in light curves of other classes \citep{Belloni2000}. Observation with ID 40703-01-35-01 belonging to class \(\beta\) \citep{Klein-Wolt2002} is assigned to class \(\theta\). This observation contains many periods of missing data, and the good time intervals contain dips similar to the class \(\theta\). The observation also contains W-shaped intervals which resemble those characteristic of class \(\theta\). Furthermore, the classifier predicts that \(\beta\) is the second most probable classification of this observation (top three predictions are \(\theta\) (37.1\%), \(\beta\) (30.4\%) and \(\delta\) (6.1\%)).\par

Two observations of class \(\chi\), 10408-01-42-00 and 40703-01-20-03, are assigned to class \(\gamma\). Both observations show significantly higher count rate and RMS then an average \(\chi\) observation, which is a possible cause of their classification. Class \(\chi\) is the second most probable class prediction for both observations; top three predictions for 10408-01-42-00 are \(\gamma\) (21.8\%), \(\chi\) (21.8\%) and \(\phi\) (14.8\%), whilst for 40703-01-20-03 they are \(\gamma\) (21.9\%), \(\chi\) (21.9\%) and \(\phi\) (14.7\%).

\begin{figure*}
	\includegraphics[width=\textwidth]{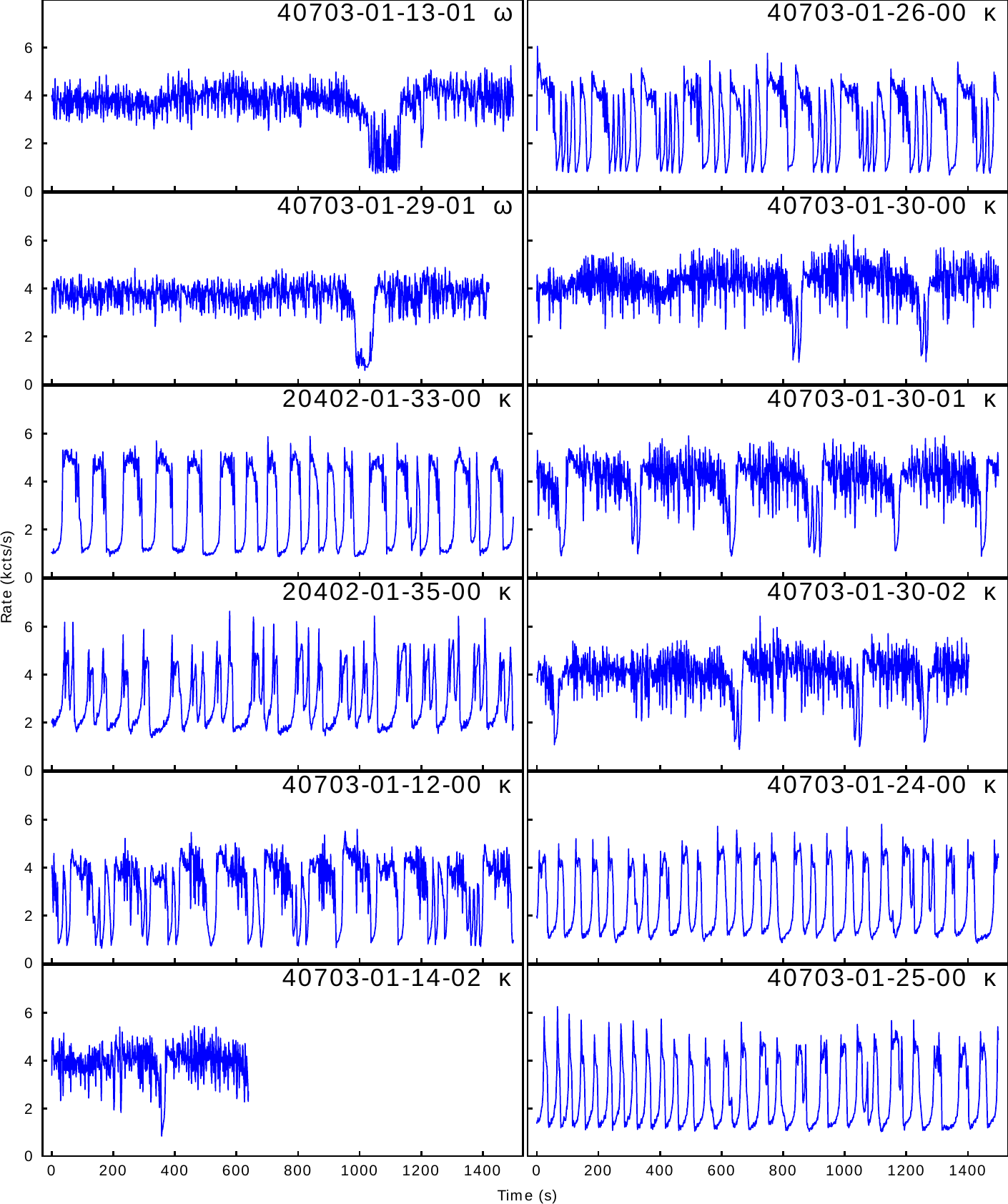}
    \caption{Light curves of \(\omega\) and \(\kappa\) observations discussed in Section~\ref{sec: kappa and omega}. Each subfigure contains first 1500 seconds of the light curve or as much as is available in case of shorter observations. Each subfigure contains the observation ID and classification from \citet{Huppenkothen2017}. Figure~\ref{fig:omega kappa light curves 3} contains light curves which require individual time axes.}
    \label{fig:omega kappa light curves 12}
\end{figure*}

Two observations of class \(\kappa\), 40703-01-24-00 and 40703-01-25-00, are assigned to class \(\rho\). Both observations show fairly regular, sharp flares, similar to those characteristic to class \(\rho\), however they are noticeably wider than the canonical \(\rho\) flares (see Figure~\ref{fig:omega kappa light curves 12} for the light curves of the two observations). Other than the similarity to \(\rho\) light curves, another factor influencing the classification of these observations is likely the heterogeneity of the light curve behaviour of observations labelled as \(\kappa\) by \citet{Klein-Wolt2002}. More details about the ambiguity of \(\kappa\) classifications is provided in Section~\ref{sec: kappa and omega}. Furthermore, class \(\kappa\) is the second most probable class prediction for both observations. Top three predictions for 40703-01-24-00 are \(\rho\) (16.0\%), \(\kappa\) (15.5\%) and \(\chi\) (14.0\%), whilst for 40703-01-25-00 they are \(\rho\) (15.7\%), \(\kappa\) (15.3\%) and \(\chi\) (13.4\%).

Observation 20402-01-36-01 belongs to class \(\lambda\) and is assigned to class \(\theta\) by the random forest classifier. This observation shows behaviour which is very characteristic of class \(\lambda\); it shows periods of flaring which alternate with low, quiet periods (see \(\lambda\) light curve in Figure~\ref{fig:light curves} for an example). The likely cause of this class assignment is the fact that only one \(\lambda\) observation is included in the training data subset. The classifier predicts that \(\lambda\) is the second most probable classification; top three predictions are \(\theta\) (30.1\%), \(\lambda\) (20.8\%) and \(\kappa\) (13.2\%).\par


Finally, observation 40703-01-29-01 belonging to class \(\omega\) is assigned to class \(\kappa\) by the classifier. This observation shows steady flux with no structured variability except a singe W-shaped dip, which is a very typical \(\omega\) behaviour (see Figure~\ref{fig:omega kappa light curves 12} for the light curve). The classifier predicts that \(\omega\) is the second most probable classification for this observation (top three predictions are \(\kappa\) (29.9\%), \(\omega\) (21.0\%) and \(\gamma\) (16.8\%)). One cause contributing to this classification is the fact that only one labelled observation of class \(\omega\) is available in the training data subset. However, another major cause becomes apparent upon inspection of \(\omega\) and \(\kappa\) observations. Many of the observations labelled as \(\kappa\) by \citet{Klein-Wolt2002} show behaviour which very strongly resembles class \(\omega\). We discuss this issue in more detail in Section~\ref{sec: kappa and omega}.

\subsection{Data-driven review of \texorpdfstring{\(\omega\)}{omega} and \texorpdfstring{\(\kappa\)}{kappa} classifications}
\label{sec: kappa and omega}

\begin{figure*}
	\includegraphics[width=\textwidth]{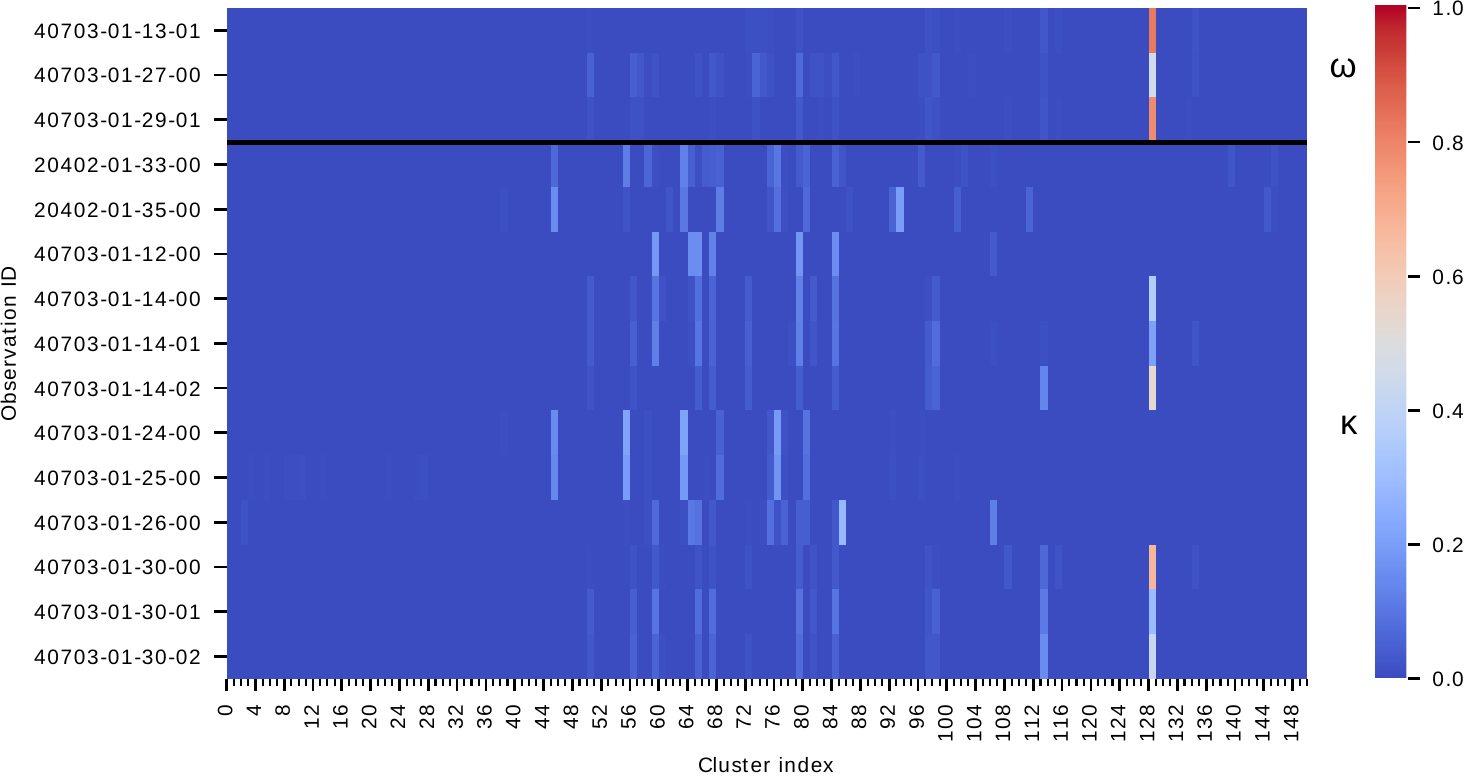}
    \caption{``Fingerprint'' representation of \(\omega\) (above black line) and \(\kappa\) (below black line) observations from \citet{Huppenkothen2017}. Clusters are merged using the method described in Section~\ref{sec: classification}, using the optimal Mahalanobis distance threshold of 3.34. Colour indicates the relative abundance of light curve segments in the corresponding cluster.}
    \label{fig:omega kappa fingerprints}
\end{figure*}

\begin{figure}
	\includegraphics[width=\columnwidth]{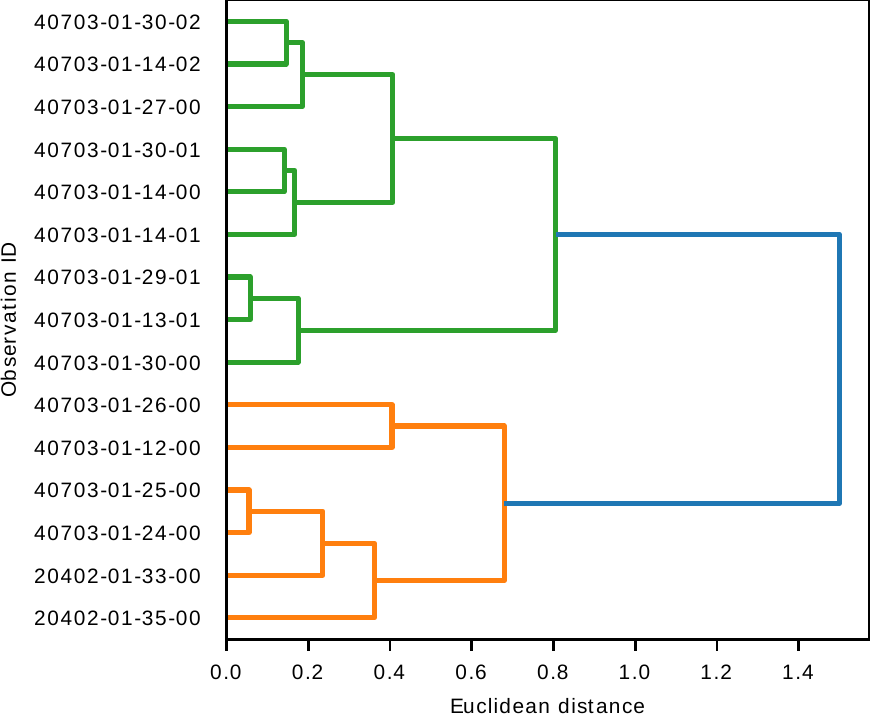}
    \caption{Dendogram resulting from the hierarchical clustering of \(\omega\) and \(\kappa\) observations based on their ``fingerprints'' shown in Figure~\ref{fig:omega kappa fingerprints}. Hierarchical clustering algorithm uses the ward method and euclidean metric. Splitting of branches of the dendogram at smaller values of Euclidean distance indicates that the observations in corresponding leaf nodes are more closely related.}
    \label{fig:omega kappa dendogram}
\end{figure}

\begin{figure}
    \includegraphics[width=\columnwidth]{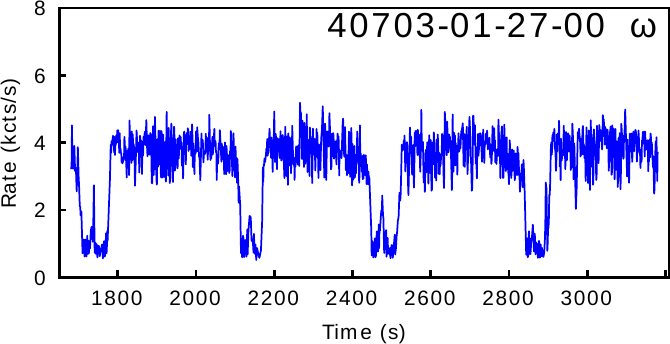}
	\includegraphics[width=\columnwidth]{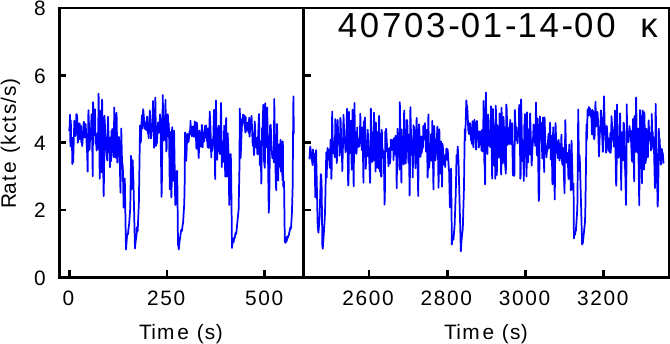}	\includegraphics[width=\columnwidth]{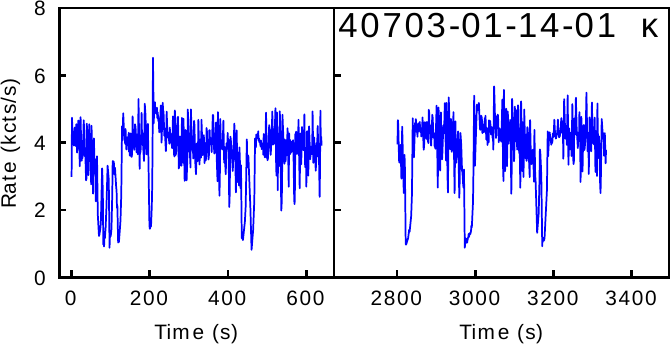}	
    \caption{Light curves of \(\omega\) and \(\kappa\) observations which are mentioned in Section~\ref{sec: classification}. Top subfigure contains only the second, longer good time interval of the observation. Middle and bottom subfigures contain two good time intervals of their respective observations, and the data gaps are removed. Each subfigure contains 1500 seconds of the light curve or as much as is available in case of shorter observations. Each subfigure shows the observation ID and classification from \citet{Huppenkothen2017}. Figure~\ref{fig:omega kappa light curves 12} contains light curves of remaining \(\omega\) and \(\kappa\) observations.}
    \label{fig:omega kappa light curves 3}
\end{figure}

\begin{table}
	\centering
	\caption{Chronological list of RXTE/PCA observations of GRS 1915+105 during periods 51288-51306 MJD and 51394-51432 MJD. Classifications from \citet{Klein-Wolt2002} (Class. K), \citet{Pahari2010} (Class. P) and \citet{Belloni2013} (Class. B) are provided.}
	\label{tab:omega kappa chronology}
	\begin{tabular}{lcccc}
		\hline
		Observation ID & Date (MJD) & Class. K & Class. P & Class. B\\
		\hline
		40703-01-12-00 $\dagger$ & 51288 & \(\kappa\) & \(\omega\) & \(\omega\)\\
		40115-01-01-00 & 51288 & -  & - & -\\
		40403-01-07-00 & 51291 & \(\omega\)  & \(\omega\) & \(\omega\)\\
		40703-01-13-00 & 51299 & \(\gamma\)  & - & -\\
		40703-01-13-01 & 51299 & \(\omega\) & \(\omega\) & \(\omega\)\\
		40115-01-02-00 & 51299 & -  & - & -\\
		40703-01-14-00 $\dagger$ & 51306 & \(\kappa\)  & \(\omega\) & - \\
		40703-01-14-01 & 51306 & \(\kappa\) & - &  - \\
		40703-01-14-02 & 51306 & \(\kappa\) & - & - \\
		\hline
		40703-01-24-00 & 51394 & \(\kappa\)  & - & -\\
		40703-01-25-00 & 51399 & \(\kappa\)  & - & -\\
		40115-01-05-00 & 51406 & - & - & -\\
		40703-01-26-00 $\dagger$ & 51407 & \(\kappa\)  & - & \(\omega\)\\
		40703-01-27-00 & 51413 & \(\omega\) & - & -\\
		40703-01-27-01 & 51413 & \(\gamma\)/\(\omega\)  & \(\omega\) & -\\
		40703-01-28-00 & 51418 & \(\omega\)  & - & \(\omega\)\\
		40703-01-28-01 & 51418 & - & - & -\\
		40703-01-28-02 & 51418 & \(\omega\)  & \(\omega\) & \(\omega\)\\
		40115-01-06-00/01 & 51423 & - & - & - \\
		40703-01-29-00 & 51426 & \(\omega\)  & \(\omega\) & \(\omega\)\\
		40703-01-29-01 & 51426 & \(\omega\)  & - & \(\omega\)\\
		40703-01-29-02 & 51426 & \(\gamma\)/\(\omega\)  & - & \(\omega\)\\
		40703-01-30-00 $\dagger$ & 51432 & \(\kappa\)  & \(\omega\) & - \\
		40703-01-30-01 & 51432 & \(\kappa\)  & - & - \\
		40703-01-30-02 & 51432 & \(\kappa\)  & - & -\\
		40703-01-30-03 & 51432 & \(\gamma\)/\(\kappa\)  & - & -\\
		\hline
	\end{tabular}
$\dagger$ Observations whose classifications seem to be inconsistent
\end{table}

Figure~\ref{fig:omega kappa fingerprints} shows the ``fingerprint'' representation of all \(\omega\) and \(\kappa\) observations from \citet{Huppenkothen2017}. There clearly exist at least two groups of \(\kappa\) observations. Six of the observations (40703-01-14-00/01/02, 40703-01-30-00/01/02) are much more similar to \(\omega\) observations than the other \(\kappa\) observations. In order to assess the similarity of presented observations, we perform hierarchical clustering of observations in the ``fingerprint'' space. We use the hierarchical clustering algorithm included in the SciPy package \citep{Virtanen2020} (see Appendix~\ref{sec: hierarchical clustering} for more details about the algorithm).\par

Figure~\ref{fig:omega kappa dendogram} shows a dendogram resulting from the clustering of ``fingerprints'' of \(\omega\) and \(\kappa\) observations (see Figure~\ref{fig:omega kappa light curves 12} and Figure~\ref{fig:omega kappa light curves 3} for their light curves). In the green branch of the dendogram, observations classified as \(\omega\) by \citet{Klein-Wolt2002} are clustered with the six \(\kappa\) observations mentioned above. These observations show semi-regular dips with one or more re-flares. Observations 40703-01-13-01/29-01/30-00 clustered in the lower green sub-branch show lower frequency of dipping than the other observations, making them more alike to the canonical \(\omega\) behaviour. Observations 40703-01-30-02/14-02/27-00 show a slightly higher dipping frequency, whilst observations 40703-01-30-01/14-00/14-01 show the highest frequency of dips in the green branch of the dendogram. Similarity between ``fingerprints'' of \(\kappa\) observations 40703-01-14-00/01/02, 40703-01-30-00/01/02, and canonical \(\omega\) observations is the result of the similarity between the light curves of those observations. Based on this similarity, we suggest that those \(\kappa\) observations should be viewed as examples of an intermediate \(\kappa\)/\(\omega\) state, with a major \(\omega\) component.\par

Observations in the orange branch of the dendogram show characteristic quasi-periodic and aperiodic flares of class \(\kappa\). However, observations 40703-01-12-00/26-00 show a mix of flares with large and small width as opposed to 40703-01-24-00/25-00 and 20402-01-33-00/35-00, show a more consistent flare profile. Moreover, the presence of flares with greater width suggests an intermediate \(\kappa\)/\(\omega\) state with a major \(\kappa\) component. Furthermore, \citet{Belloni2013} indicate that observations 40703-01-12-00/26-00 belong to class \(\omega\), whilst both \citet{Pahari2010} and \citet{Belloni2013} indicate that observations 40703-01-14-00/30-00 belong to class \(\omega\), which shows that ambiguity of \(\kappa\) and \(\omega\) classification exists in the literature. \par

Table~\ref{tab:omega kappa chronology} shows a chronological list of observations captured in the periods when the source was showing \(\kappa\) and \(\omega\) behaviour, along with the classifications from \citet{Klein-Wolt2002}, \citet{Pahari2010} and \citet{Belloni2013}. Observations of class \(\kappa\) and \(\omega\) were observed in close succession, which supports the notion of intermediate states between those classes. Based on the classification of observations in the two periods shown in Table~\ref{tab:omega kappa chronology}, it seems that the source tends to transition between classes in the order \(\kappa \rightarrow \omega \rightarrow \kappa\).\par

However, clustering results of the ``fingerprint'' representation should be interpreted with caution due to the geometry of the data sampling space (see Section~\ref{sec:conclusions} for more details).

\subsection{Classification of 1024 second segments}
\label{sec: comparison to Huppenkothen+2017}

We conduct an additional classification experiment with the goal of making it as comparable as possible to the work of \citet{Huppenkothen2017}. The purpose of our paper is primarily to introduce an unsupervised method of light curve data characterisation, whilst the primary objective of \citet{Huppenkothen2017} was to study GRS 1915+105, however both papers use machine learning methods to perform classification on this source, so in spite of the differences, we believe that this experiment is informative. \par

We segment all the available RXTE/PCA observations of GRS 1915+105 in Standard-1 format to 1024 second segments and use the stride length of 256 seconds, yielding the total of 11028 segments, 2141 of which have human-assigned labels. We split this data between the training, validation and testing data subsets in the ratio of approximately 50:25:25, ensuring that segments created from a single observation are not split between data subsets. Data is split in a stratified manner; at least one observation of each class is randomly assigned to each subset, and the remaining observations are distributed according to the ratio of 50:25:25. This results in 982, 592 and 566 segments in the training, validation and testing subsets respectively. Further 6573 and 2314 segments of observations with no human-assigned labels are also assigned to the training and validation sets respectively. \par

In order to make the segments compatible with the LSTM-VAE network discussed in Section~\ref{sec:Encoding shape features}, we further subdivide the 1024 second segments into 64 non-overlapping segments of 16 seconds (which at the cadence of 0.125 seconds comprise 128 data points, as required by the network). We train the network using the training subset and evaluate the performance of the network after each epoch of training using the validation subset of data. We use the Adam optimiser with the \texttt{clipvalue} argument set to 0.5 to train the network. Training is stopped after 269 epochs, when the validation loss stopped improving for 50 consecutive epochs. We use the resulting network to encode the light curve segments and create 20 SFoS features. \par

We fit a Gaussian mixture model with 250 components to the SIFoS features of the training and validation subsets of data. We then perform a grid search of hyper-parameters listed in Table~\ref{tab:random forest hyperparameters}, where for each combination of hyper-parameters we train a random forest classifier on the training set and we evaluate the prediction on the validation set. We find that the highest weighted F1 and accuracy performance scores of 0.834 and 0.851 respectively are produced using the merge distance of 3.798, criterion of ``entropy'', and max\_depth of 5.\par

\begin{figure}
	\includegraphics[width=\columnwidth]{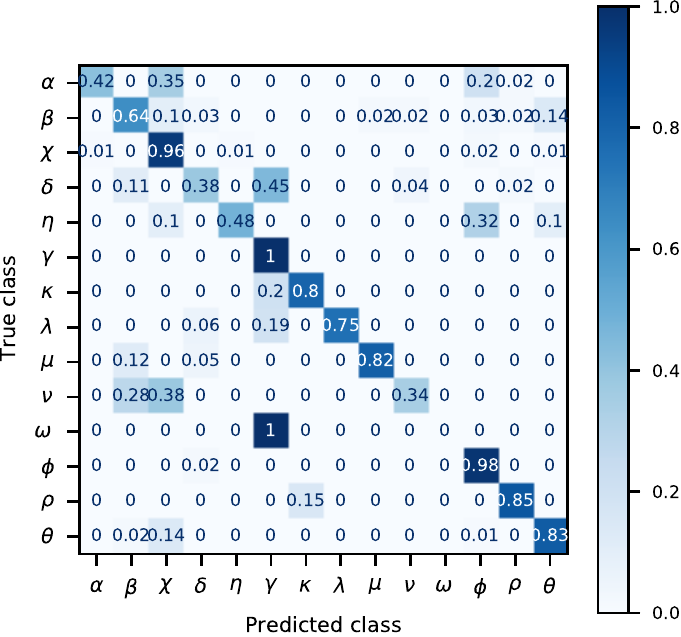}
    \caption{A row-wise normalised confusion matrix showing classification results for light curve segments of 1024 seconds. The numbers in each row are divided by the sum of numbers in that row.}
    \label{fig: Huppenkothen comparison confusion matrix}
\end{figure}

We test the classifier with optimal hyper-parameters on combined validation and testing data subsets, resulting in weighted F1 and accuracy performance scores of 0.811 and 0.822 respectively. The accuracy of this classification is ca. 10 percentage points lower than that of \citet{Huppenkothen2017}. Their method is simpler and more accurate, but it requires a larger variety of data features, i.e. power-spectral and hardness ratio features. Therefore, depending on the available features, it might be the preferable method for the task of classification of 1024 second segments. Including those features in our method could result in a performance improvement, which is a possible avenue of future investigation.

The possible avenues for improvement of both methods include the training on a larger amount of labelled data, as well as the reduction of ambiguity in the labelled data. Ambiguity is caused by the fact that the classes of the Belloni et al. system are not fully exhaustive and mutually exclusive. This is discussed further in Section~\ref{sec:conclusions}. \par

Figure~\ref{fig: Huppenkothen comparison confusion matrix} shows the confusion matrix of the predicted labels against the human-assigned labels. Even though most of the segments are classified in agreement with the human-assigned labels, the drop in performance is significant when compared to the other experiments shown in this paper (Section~\ref{sec: classification} and Appendix~\ref{sec: validation classifications}). In particular, all of the \(\omega\) segments are classified as \(\gamma\), which is most likely caused by 
the noise in the data. The data with cadence of 0.125 second contains a significant amount of random noise, and the segmentation of data to the length of only 16 seconds followed by standardisation, results in segments which are seemingly featureless, with occasional peaks and troughs beyond the noise level. The characteristic dips of class \(\omega\) were not preserved in this representation of data. In conclusion, re-binning of data to cadence longer than 0.125 seconds prior to LSTM-VAE training is likely to be beneficial for the performance of our method. Additionally, when light curves are segmented to 1024 seconds prior to classification, we cannot exploit the main advantage of the ``fingerprint" representation, which aggregates features of long light curves. Light curves of classes whose cycles of variability evolve over longer time than 1024 seconds can yield segments which appear significantly different, depending on the phase of a variability cycle. \par

\section{Conclusions}
\label{sec:conclusions}

We introduce a data-driven method of light curve feature extraction and test the utility of resulting features by conducting a set of supervised multi-class classification experiments, using a set of human-labelled observations. Light curve classification of data in 1 second resolution resulting in a mean weighted F1 score of \(0.878\) suggests that the proposed method of unsupervised feature extraction is capable of producing features which represent light curve data in a meaningful way. \par

In regard to potential future work on the data set of GRS 1915+105 X-ray light curves, a supervised approach could be taken to study the link between the observed light curve patterns and underlying physics. \citet{Huppenkothen2017} used a classification scheme of GRS 1915+105 states into stochastic and chaotic processes \citep{Harikrishnan2011} in order to leverage their machine learning classification algorithm in the study of the long term evolution of accretion properties of the source, and similar studies could be complemented using our method of feature extraction.\par

Another possible application of the method is the unsupervised exploration of the data space. The shape of the latent data manifold encodes information about modes of activity of the source and its evolution between them. The analysis of correlation between SIFoS and various power-spectral and energy-spectral features could be performed in order to study the links between latent variables, the occurrence of different variability patterns and the physically interpretable parameters. For example, broad-band power spectral shape could help to link variability patterns with the geometry of the emitting regions \citep{Heil2015}, presence of high frequency quasi-periodic oscillations could link the patterns with the orbital motion of matter in the accretion disc \citep[see][for a review]{Motta2016}, and radio flux could link them with the emission of relativistic jets \citep{Belloni2011}.\par

Work presented in this paper shows that dimensionality reduction of the data set, followed by clustering of observations in this reduced space could be a way to derive a set of classes of source behaviour, which avoids the biases of human characterisation and annotation of data. Furthermore, we find that Gaussian component merging based on the Mahalanobis distance between them can help to reduce the problem of cluster degeneracy caused by the limitations of GMM, which requires multiple components to follow curved data manifolds.\par

The Belloni et al. system of classification discussed in this work is not comprehensive, and to some degree, it is arbitrary. As \citet{Belloni2000} point out, their classification system was not intended to exhaustively list mutually exclusive modes of behaviour. This creates a problem for any classification effort that is based on this system. A smaller number of classes could be chosen, because some classes show behaviour that arguably lies on the same continuum. One example of such continuum could be followed by classes \(\lambda\) and \(\kappa\), which show similar behaviour at slightly different time scales \citep{Belloni2000}. Furthermore, our work shows that classes \(\kappa\) and \(\omega\) can show very similar light curve behaviour and possibly lie on a continuum as well. A larger number of classes could just as well be required, as it is possible that there exist additional patterns of behaviour that have not been characterised yet (similarly to classes \(\omega\) and \(\xi\) (a.k.a. \(\eta\)) which were added to the Belloni et al. system later than the original 12). Transitions between classes are observed, so ambiguity in classification of observations cannot be avoided.\par

Taking these considerations into account, it is very difficult to ensure the accuracy of classification for large sets of unknown data, which cannot be standardised to adhere to the assumed classification system. Performance of any supervised classification algorithm greatly depends on the definition of the classes of observations, and any ambiguity is going to affect this performance. Therefore, clustering of the data set can help in deriving a data-driven set of observation classes, which helps to avoid the biases of human characterisation and annotation of data. \par

However, further work in this direction will need to address the problem of clustering of compositional data. The ``fingerprint'' representation uses vectors of fixed size, whose values sum up to a positive constant. This is a type of compositional data, and as such it is constrained to the geometry of a simplex \citep{Aitchison2005}, which means that results of clustering of raw compositional data are often unreliable. Clustering methods applied to compositional data should take into account the need of prior transformation of the data into an unbounded, Euclidean space. This issue is an open area of research, and it is further complicated by the presence of a large number of zero values in the ``fingerprint'' compositional data \citep{Aitchison2000, Martin-Fernandez2003}. Study of the appropriate transformation methods for compositional data goes beyond the scope of our work, therefore results of ``fingerprint'' clustering should be interpreted with caution. \par

Since the proposed feature extraction method is easily generalisable to different types of time series data, there exists a range of possible applications for the proposed feature extraction pipeline. A similar type of analysis is possible for sources other than GRS 1915+105, and in principle any energy band of light curves.\par

Fingerprint representation could be used to test whether the phenomenology observed in GRS 1915+105 is also present in other sources. For example, similarity in the variability of IGR J17091-3624 \citep{Court2017}, and the Rapid Burster, MXB  1730-335 \citep{Bagnoli2015} could point to accretion physics which are independent of the nature of the accretor.\par

\citet{Pattnaik2021} attempted machine learning classification of compact objects in low mass X-ray binaries based on energy-spectral features, and found that fairly accurate classification was possible. The presence of a black hole or neutron star in the binary system can have a significant impact on the physical interpretation of the observed phenomenology. Classification of compact objects using the fingerprint representation of X-ray variability patterns is a possible subject of future work. However, care must be taken to select sources in the hard state, which show structured X-ray variability.\par

Derivation of fingerprints which represent the set of light curve patterns observed in a fixed amount of time could be the basis of a live monitoring system, which would alert the user about changes in the behaviour of the source. This could involve classifying observations using a known system of classes, but it could also involve the task of outlier detection, where position of an observation would be tracked within the encoded feature space. Observations producing feature vectors which fall in sparse regions of the feature space would indicate an anomaly.\par

The main requirement of the proposed feature extraction method is that the light curves must be evenly sampled in time. This requirement is satisfied by data similar in nature to the pointed observations of RXTE/PCA. This includes data captured by the X-ray Telescope aboard the Neil Gehrels Swift Observatory (Swift) \citep{Burrows2005} or X-ray Timing Instrument aboard the Neutron Star Interior Composition Explorer (NICER) \citep{Gendreau2016} etc. High-speed optical telescopes ULTRACAM \citep{Dhillon2007} and HiPERCAM \citep{Dhillon2016} also produce light curves which are evenly sampled over the time of an exposure, and could be analysed using the method we propose. However, the amount of data produced by ULTRACAM and HiPERCAM is not large enough to justify the use of machine learning analysis. The Optical Timing Camera (OPTICAM) \citep{Castro2019} will produce a larger amount of data of similar nature, and the proposed method of feature extraction could be appropriate for their analysis.\par

The proposed method could be used to characterise long-term light curves captured by all-sky X-ray surveys, like those performed with the Gas Slit Camera aboard MAXI \citep{Matsuoka2009}, the Burst Alert Telescope aboard Swift \citet{Gehrels2004} or the All Sky Monitor aboard RXTE, provided that light curves have regular time bins. Interpolation could also be performed if needed.\par

\citet{Pursiainen2020} interpolated ca. 30,000 light curves of the Dark Energy Survey Supernova Programme using Gaussian Processes, and increased the cadence from the average of 7 days to a constant 0.5 day cadence \citep[see also][]{Wiseman2020}. Light curves generated by several ground-based surveys could be made compatible with our feature extraction method using such interpolation techniques. Those surveys include LSST, ZTF and Asteroid Terrestrial-impact Last Alert System (ATLAS) \citep{Heinze2018}. Optimisation of the type and size of kernel used in Gaussian Processes interpolation would need to be performed by the user prior to feature extraction, and the choice of parameters would depend on the nature of data and the scientific goal.\par 

In addition to the requirement of having even sampling in time, light curves should not have many gaps which cannot be interpolated over, in order to be suitable for analysis using the proposed feature extraction method. During the light curve segmentation stage, segments are extracted using a moving window method, and segments which straddle over data gaps are discarded. Therefore, light curves must have few gaps to allow for a choice of segment size which is large enough to encompass time-scales which are relevant for the variability of the analysed data.\par

There are several limitations to this work. Figure~\ref{fig:stat boxplot} reveals that classes \(\phi\) and \(\chi\) are similar in terms of IFoS. The only significant differences are (1) that mean count rate values of segments of class \(\chi\) span a larger range than the segments of class \(\phi\), and (2) that segments of class \(\chi\) can have more positive outliers in terms of kurtosis. Figure~\ref{fig:SFoS and IFoS UMAP} shows that the combination of IFoS contains enough information to distinguish many cases of those classes, but significant areas of overlap still exist. The two classes are best distinguished based on the hardness of their colour-colour diagrams \citep{Belloni2000}. Since both classes of observations show no structured X-ray variability patterns, and their IFoS distributions largely overlap, it should be noted that classification attempts based on features presented in this work could disagree with classification of \citet{Belloni2000}. The proposed method is only able to capture time series patterns, therefore it would not be able to differentiate between observations which differ only in terms of energy spectra. Nevertheless, users of the method could choose to supplement SIFoS with additional features, dependent on their individual use case.\par 

Another set of limitations involves data pre-processing. The re-binning of light curve data to 1 and 4 second resolution was a compromise between computational tractability and descriptiveness of GRS 1915+105 variability. Even though the shortest timescale of significant X-ray flux changes in this source is ca. 5 seconds \citep{Nayakshin2000}, binning inevitably led to a loss of some of the fast variability information. Furthermore, the choice of light curve segment length was influenced by computational constrains, but also informed by the previous work and knowledge of the time scales of light curve patterns in GRS 1915+105. There is a fairly large degree of tolerance for the choice of these parameters, but some knowledge of the time scales of interest was required. \par

We did not fine-tune the method to study any particular subset of variability classes within the source, however users interested in a more specialised study of a particular scientific problem could choose to analyse light curves binned to smaller time bins to improve the modelling of very narrow features. For example, \cite{Massaro2020-II} modelled the light curves of GRS 1915+105 using a system of ordinary differential equation, and successfully reproduced the sharp peaks present at the beginning of the \(\kappa\) class bursts. Reconstruction of light curve segments binned to 1 second resolution results in some smearing of such fine features (see Figure~\ref{fig:segment reconstructions 1s}), so the study of this particular feature would benefit from higher time resolution.
However, the light curves of GRS 1915+105 contain a significant amount of noise, and increasing the temporal resolution of light curves will result in the increased amount of noise in the training data set, which could affect the ability of the model to identify patterns within the data.\par

For cases where relevant time scales are difficult to predict, multiple ``fingerprints'' could be derived from data in a range of temporal resolutions, and subsequently concatenated to create a single feature vector. However such an approach would increase the noise in feature vectors, so further work is required to test this.\par

Our main assumption was that light curve segments exhibiting similar type of variability patterns and count rate distribution had similar values of SFoS and IFoS. Segments with more similar values of SIFoS were consequently separated by smaller distances within the SIFoS feature space. Therefore, it was expected that Gaussian clusters contained homogeneous subsets of light curve segments, which were more similar to each other than to segments found in other Gaussian clusters. Inspection of Gaussian cluster populations revealed that this assumption was justified to a large degree, however fitting a GMM with a larger number of Gaussian components would provide a greater precision of density estimation of data in the feature space, and hence reduce the risk of non-homogeneity of created clusters. On the other hand, such an approach would likely aggravate the issue of pattern degeneracy in the resulting set of light curve patterns, which would potentially lead to more complex ``fingerprints'' and ultimately to a more noisy feature set. \par

It should be noted that GMM is not inherently a clustering algorithm but a density estimation algorithm. Following preliminary experiments with alternative clustering algorithms we find the approach of merging Gaussian components of GMM to be the most straightforward method of clustering for this data set. However it is possible that the issue of cluster degeneracy could be circumvented if a better-suited clustering method was found. This is an important area of future work.

Merging of clusters based on the Mahalanobis distance can help to reduce the number of degenerate features, but in this work we rely on the classified subset of data to find the optimal distance threshold. In cases where data classification is not possible, unsupervised methods of evaluating clustering performance could be used instead, for example Silhouette Coefficient, Calinski-Harabasz Index and Davies-Bouldin Index included in the scikit-learn library \citep{Pedragosa2011}, however computational cost of such an approach might be higher. \par

\section*{Data availability}

The data underlying this article are publicly available in the High Energy Astrophysics Science Archive Research Center (HEASARC) at \url{https://heasarc.gsfc.nasa.gov/db-perl/W3Browse/w3browse.pl}, and in the repository of the article of \citet{Huppenkothen2017} at \url{https://github.com/dhuppenkothen/BlackHoleML}.

\section*{Acknowledgements}
We thank Tomaso Belloni for sharing his work on GRS 1915+105. We also thank Kevin Alabarta, Mehtap Ozbey Arabaci, John Paice, Philip Wiseman, Miika Pursiainen and the data scientists of HAL24K Labs for useful discussion and advice. 
DA and ABH acknowledge support from the Royal Society.




\bibliographystyle{mnras}
\bibliography{references} 

\begin{thebibliography}{}
\makeatletter
\relax
\def\mn@urlcharsother{\let\do\@makeother \do\$\do\&\do\#\do\^\do\_\do\%\do\~}
\def\mn@doi{\begingroup\mn@urlcharsother \@ifnextchar [ {\mn@doi@}
  {\mn@doi@[]}}
\def\mn@doi@[#1]#2{\def\@tempa{#1}\ifx\@tempa\@empty \href
  {http://dx.doi.org/#2} {doi:#2}\else \href {http://dx.doi.org/#2} {#1}\fi
  \endgroup}
\def\mn@eprint#1#2{\mn@eprint@#1:#2::\@nil}
\def\mn@eprint@arXiv#1{\href {http://arxiv.org/abs/#1} {{\tt arXiv:#1}}}
\def\mn@eprint@dblp#1{\href {http://dblp.uni-trier.de/rec/bibtex/#1.xml}
  {dblp:#1}}
\def\mn@eprint@#1:#2:#3:#4\@nil{\def\@tempa {#1}\def\@tempb {#2}\def\@tempc
  {#3}\ifx \@tempc \@empty \let \@tempc \@tempb \let \@tempb \@tempa \fi \ifx
  \@tempb \@empty \def\@tempb {arXiv}\fi \@ifundefined
  {mn@eprint@\@tempb}{\@tempb:\@tempc}{\expandafter \expandafter \csname
  mn@eprint@\@tempb\endcsname \expandafter{\@tempc}}}

\bibitem[\protect\citeauthoryear{Aitchison \& Egozcue}{Aitchison \&
  Egozcue}{2005}]{Aitchison2005}
Aitchison J.,  Egozcue J.~J.,  2005, \mn@doi [Mathematical Geology]
  {10.1007/s11004-005-7383-7}, 37, 829

\bibitem[\protect\citeauthoryear{Aitchison, Barcel{\'{o}}-Vidal,
  Mart{\'{i}}n-Fern{\'{a}}ndez  \& Pawlowsky-Glahn}{Aitchison
  et~al.}{2000}]{Aitchison2000}
Aitchison J.,  Barcel{\'{o}}-Vidal C.,  Mart{\'{i}}n-Fern{\'{a}}ndez J.~A.,
  Pawlowsky-Glahn V.,  2000, \mn@doi [Mathematical Geology]
  {https://doi.org/10.1023/A:1007529726302}, 32, 271

\bibitem[\protect\citeauthoryear{Altamirano et~al.,}{Altamirano
  et~al.}{2011}]{Altamirano2011}
Altamirano D.,  et~al., 2011, \mn@doi [The Astrophysical Journal Letters]
  {https://doi.org/10.1088/2041-8205/742/2/L17}, 742

\bibitem[\protect\citeauthoryear{Ankerst, Breunig, Kriegel  \& Sander}{Ankerst
  et~al.}{1999}]{Ankerst1999}
Ankerst M.,  Breunig M.~M.,  Kriegel H.-p.,   Sander J.,  1999, ACM SIGMOD
  Record, 28, 49

\bibitem[\protect\citeauthoryear{Armstrong et~al.,}{Armstrong
  et~al.}{2015}]{Armstrong2015}
Armstrong D.~J.,  et~al., 2015, \mn@doi [Monthly Notices of the Royal
  Astronomical Society] {10.1093/mnras/stv2836}, 456, 2260

\bibitem[\protect\citeauthoryear{Bagnoli \& In't~Zand}{Bagnoli \&
  In't~Zand}{2015}]{Bagnoli2015}
Bagnoli T.,  In't~Zand J.~J.,  2015, \mn@doi [Monthly Notices of the Royal
  Astronomical Society: Letters] {10.1093/mnrasl/slv045}, 450, L52

\bibitem[\protect\citeauthoryear{Becker, Pichara, Catelan, Protopapas, Aguirre
  \& Nikzat}{Becker et~al.}{2020}]{Becker2020}
Becker I.,  Pichara K.,  Catelan M.,  Protopapas P.,  Aguirre C.,   Nikzat F.,
  2020, \mn@doi [Monthly Notices of the Royal Astronomical Society]
  {10.1093/mnras/staa350}, 493, 2981

\bibitem[\protect\citeauthoryear{Bellm}{Bellm}{2014}]{Bellm2014}
Bellm E.~C.,  2014, in The Third Hot-wiring the Transient Universe Workshop. pp
  27--33, \url {http://arxiv.org/abs/1410.8185}

\bibitem[\protect\citeauthoryear{Belloni}{Belloni}{2001}]{Belloni2001}
Belloni T.,  2001, in , Vol.~567, The Neutron Star-Black Hole Connection.
Kluwer Academic Publishers, pp 295--300

\bibitem[\protect\citeauthoryear{Belloni \& Altamirano}{Belloni \&
  Altamirano}{2013}]{Belloni2013}
Belloni T.~M.,  Altamirano D.,  2013, \mn@doi [Monthly Notices of the Royal
  Astronomical Society] {10.1093/mnras/stt500}, 432, 10

\bibitem[\protect\citeauthoryear{Belloni, M{\'{e}}ndez, King, Van Der~Klis  \&
  Van~Paradijs}{Belloni et~al.}{1997a}]{Belloni1997a}
Belloni T.,  M{\'{e}}ndez M.,  King A.~R.,  Van Der~Klis M.,   Van~Paradijs J.,
   1997a, \mn@doi [The Astrophysical Journal Letters]
  {https://doi.org/10.1086/310595}, 479, 145

\bibitem[\protect\citeauthoryear{Belloni, M{\'{e}}ndez, King, Van Der~Klis  \&
  Van~Paradijs}{Belloni et~al.}{1997b}]{Belloni1997b}
Belloni T.,  M{\'{e}}ndez M.,  King A.~R.,  Van Der~Klis M.,   Van~Paradijs J.,
   1997b, \mn@doi [The Astrophysical Journal Letters]
  {https://doi.org/10.1086/310944}, 488, 109

\bibitem[\protect\citeauthoryear{Belloni, Klein-Wolt, Mendez, van~der Klis  \&
  van Paradijs}{Belloni et~al.}{2000}]{Belloni2000}
Belloni T.,  Klein-Wolt M.,  Mendez M.,  van~der Klis M.,   van Paradijs J.,
  2000, Astronomy and Astrophysics, 355, 271

\bibitem[\protect\citeauthoryear{Belloni, Motta  \& Mu{\~{n}}oz-Darias}{Belloni
  et~al.}{2011}]{Belloni2011}
Belloni T.~M.,  Motta S.~E.,   Mu{\~{n}}oz-Darias T.,  2011, Bull. Astr. Soc.
  India, 39, 409

\bibitem[\protect\citeauthoryear{Benkabou, Benabdeslem  \& Canitia}{Benkabou
  et~al.}{2018}]{Benkabou2018}
Benkabou S.-E.,  Benabdeslem K.,   Canitia B.,  2018, \mn@doi [Knowledge and
  Information Systems] {10.1007/s10115-017-1067-8}, 54, 463

\bibitem[\protect\citeauthoryear{Breiman}{Breiman}{2001}]{Breiman2001}
Breiman L.,  2001, Machine Learning, 45, 5

\bibitem[\protect\citeauthoryear{Burrows et~al.,}{Burrows
  et~al.}{2005}]{Burrows2005}
Burrows D.~N.,  et~al., 2005, \mn@doi [Space Science Reviews]
  {10.1007/s11214-005-5097-2}, 120, 165

\bibitem[\protect\citeauthoryear{Capitanio et~al.,}{Capitanio
  et~al.}{2006}]{Capitanio2006}
Capitanio F.,  et~al., 2006, The Astrophysical Journal, 643, 376

\bibitem[\protect\citeauthoryear{Castro-Tirado, Brandt, Lund, Lapshov, Sunyaev,
  Shlyapnikov, Guziy  \& Pavlenko}{Castro-Tirado
  et~al.}{1994}]{Castro-Tirado1994}
Castro-Tirado A.~J.,  Brandt S.,  Lund N.,  Lapshov I.,  Sunyaev R.~A.,
  Shlyapnikov A.~A.,  Guziy S.,   Pavlenko E.~P.,  1994, \mn@doi [Astrophysical
  Journal Supplement] {10.1086/191998}, 92, 469

\bibitem[\protect\citeauthoryear{Castro et~al.,}{Castro
  et~al.}{2019}]{Castro2019}
Castro A.,  et~al., 2019, \mn@doi [Revista Mexicana de Astronomia y
  Astrofisica] {10.22201/ia.01851101p.2019.55.02.20}, 55, 363

\bibitem[\protect\citeauthoryear{Charnock \& Moss}{Charnock \&
  Moss}{2017}]{Charnock2017}
Charnock T.,  Moss A.,  2017, \mn@doi [The Astrophysical Journal]
  {10.3847/2041-8213/aa603d}, 837, L28

\bibitem[\protect\citeauthoryear{Chetlur, Woolley, Vandermersch, Cohen, Tran,
  Catanzaro  \& Shelhamer}{Chetlur et~al.}{2014}]{Chetlur2014}
Chetlur S.,  Woolley C.,  Vandermersch P.,  Cohen J.,  Tran J.,  Catanzaro B.,
   Shelhamer E.,  2014, {cuDNN: Efficient Primitives for Deep Learning}, \url
  {http://arxiv.org/abs/1410.0759}

\bibitem[\protect\citeauthoryear{Chollet}{Chollet}{2015}]{chollet2015keras}
Chollet F.,  2015, {Keras}, \url {https://keras.io}

\bibitem[\protect\citeauthoryear{Court, Altamirano, Pereyra, Boon, Yamaoka,
  Belloni, Wijnands  \& Pahari}{Court et~al.}{2017}]{Court2017}
Court J.~M.,  Altamirano D.,  Pereyra M.,  Boon C.~M.,  Yamaoka K.,  Belloni
  T.,  Wijnands R.,   Pahari M.,  2017, \mn@doi [Monthly Notices of the Royal
  Astronomical Society] {10.1093/mnras/stx773}, 468, 4748

\bibitem[\protect\citeauthoryear{Dempster, Laird  \& Rubin}{Dempster
  et~al.}{1977}]{Dempster1977}
Dempster A.~P.,  Laird N.~M.,   Rubin D.~B.,  1977, Journal of the Royal
  Statistical Society, Series B, 39, 1

\bibitem[\protect\citeauthoryear{Dhillon et~al.,}{Dhillon
  et~al.}{2007}]{Dhillon2007}
Dhillon V.~S.,  et~al., 2007, \mn@doi [Monthly Notices of the Royal
  Astronomical Society] {10.1111/j.1365-2966.2007.11881.x}, 378, 825

\bibitem[\protect\citeauthoryear{Dhillon et~al.,}{Dhillon
  et~al.}{2016}]{Dhillon2016}
Dhillon V.~S.,  et~al., 2016, in Proceedings of the SPIE. pp 99080Y--undefined,
  \mn@doi{10.1117/12.2229055}, \url {http://arxiv.org/abs/1606.09214
  http://dx.doi.org/10.1117/12.2229055}

\bibitem[\protect\citeauthoryear{Ester, Kriegel, Sander  \& Xu}{Ester
  et~al.}{1996}]{Ester1996}
Ester M.,  Kriegel H.-P.,  Sander J.,   Xu X.,  1996, in Proceedings of the 2nd
  International Conference on Knowledge Discovery and Data Mining. pp 226--231,
  \url {www.aaai.org}

\bibitem[\protect\citeauthoryear{Fender \& Belloni}{Fender \&
  Belloni}{2004}]{Fender2004}
Fender R.,  Belloni T.,  2004, \mn@doi [Annual Review of Astronomy and
  Astrophysics] {10.1146/annurev.astro.42.053102.134031}, 42, 317

\bibitem[\protect\citeauthoryear{Gehrels et~al.,}{Gehrels
  et~al.}{2004}]{Gehrels2004}
Gehrels N.,  et~al., 2004, The Astrophysical Journal, 611, 1005

\bibitem[\protect\citeauthoryear{Gendreau et~al.,}{Gendreau
  et~al.}{2016}]{Gendreau2016}
Gendreau K.~C.,  et~al., 2016, in Space Telescopes and Instrumentation 2016:
  Ultraviolet to Gamma Ray. SPIE, p. 99051H, \mn@doi{10.1117/12.2231304}

\bibitem[\protect\citeauthoryear{Glasser, Odell  \& Seufert}{Glasser
  et~al.}{1994}]{Glasser1994}
Glasser C.~A.,  Odell C.~E.,   Seufert S.~E.,  1994, \mn@doi [IEEE Transactions
  on Nuclear Science] {10.1109/23.322911}, 41, 1343

\bibitem[\protect\citeauthoryear{Hannikainen et~al.,}{Hannikainen
  et~al.}{2003}]{Hannikainen2003}
Hannikainen D.~C.,  et~al., 2003, \mn@doi [Astronomy {\&} Astrophysics]
  {10.1051/0004-6361:20031444}, 411, 415

\bibitem[\protect\citeauthoryear{Hannikainen et~al.,}{Hannikainen
  et~al.}{2005}]{Hannikainen2005}
Hannikainen D.~C.,  et~al., 2005, \mn@doi [Astronomy {\&} Astrophysics]
  {10.1051/0004-6361:20042250}, 435, 995

\bibitem[\protect\citeauthoryear{Harikrishnan, Misra  \& Ambika}{Harikrishnan
  et~al.}{2011}]{Harikrishnan2011}
Harikrishnan K.~P.,  Misra R.,   Ambika G.,  2011, \mn@doi [Research in
  Astronomy and Astrophysics] {10.1088/1674-4527/11/1/004}, 11, 71

\bibitem[\protect\citeauthoryear{Heil, Uttley  \& Klein-Wolt}{Heil
  et~al.}{2015}]{Heil2015}
Heil L.~M.,  Uttley P.,   Klein-Wolt M.,  2015, \mn@doi [Monthly Notices of the
  Royal Astronomical Society] {10.1093/mnras/stv240}, 448, 3348

\bibitem[\protect\citeauthoryear{Heinze et~al.,}{Heinze
  et~al.}{2018}]{Heinze2018}
Heinze A.~N.,  et~al., 2018, \mn@doi [The Astronomical Journal]
  {10.3847/1538-3881/aae47f}, 156

\bibitem[\protect\citeauthoryear{Hinton}{Hinton}{2013}]{HintonLecture}
Hinton G.,  2013, {Non-linear dimensionality reduction}, \url
  {https://www.cs.toronto.edu/~hinton/csc2535/notes/lec11new.pdf}

\bibitem[\protect\citeauthoryear{Hochreiter \& Urgen~Schmidhuber}{Hochreiter \&
  Urgen~Schmidhuber}{1997}]{Hochreiter1997}
Hochreiter S.,  Urgen~Schmidhuber J.,  1997, \mn@doi [Neural Computation]
  {https://doi.org/10.1162/neco.1997.9.8.1735}, 9, 1735

\bibitem[\protect\citeauthoryear{Huppenkothen, Heil, Hogg  \&
  Mueller}{Huppenkothen et~al.}{2017}]{Huppenkothen2017}
Huppenkothen D.,  Heil L.~M.,  Hogg D.~W.,   Mueller A.,  2017, \mn@doi
  [Monthly Notices of the Royal Astronomical Society] {10.1093/mnras/stw3190},
  466, 2364

\bibitem[\protect\citeauthoryear{Hyndman, Wang  \& Laptev}{Hyndman
  et~al.}{2015}]{Hyndman2016}
Hyndman R.~J.,  Wang E.,   Laptev N.,  2015, in 2015 IEEE International
  Conference on Data Mining Workshop (ICDMW). pp 1616--1619,
  \mn@doi{10.1109/ICDMW.2015.104}

\bibitem[\protect\citeauthoryear{Ismail~Fawaz, Forestier, Weber, Idoumghar  \&
  Muller}{Ismail~Fawaz et~al.}{2019}]{IsmailFawaz2019}
Ismail~Fawaz H.,  Forestier G.,  Weber J.,  Idoumghar L.,   Muller P.~A.,
  2019, \mn@doi [Data Mining and Knowledge Discovery]
  {10.1007/s10618-019-00619-1}, 33, 917

\bibitem[\protect\citeauthoryear{Ivezic et~al.,}{Ivezic
  et~al.}{2019}]{Ivezic2019}
Ivezic Z.,  et~al., 2019, \mn@doi [The Astrophysical Journal]
  {10.3847/1538-4357/ab042c}, 873, 111

\bibitem[\protect\citeauthoryear{Kingma \& Ba}{Kingma \& Ba}{2015}]{Kingma2015}
Kingma D.~P.,  Ba J.,  2015, in 3rd International Conference on Learning
  Representations. \url {http://arxiv.org/abs/1412.6980}

\bibitem[\protect\citeauthoryear{Kingma \& Welling}{Kingma \&
  Welling}{2014}]{Kingma2014}
Kingma D.~P.,  Welling M.,  2014, in Proceedings of the 2nd International
  Conference on Learning Representations (ICLR). \url
  {http://arxiv.org/abs/1312.6114}

\bibitem[\protect\citeauthoryear{Klein-Wolt, Fender, Pooley, Belloni, Migliari,
  Morgan  \& van~der Klis}{Klein-Wolt et~al.}{2002}]{Klein-Wolt2002}
Klein-Wolt M.,  Fender R.~P.,  Pooley G.~G.,  Belloni T.,  Migliari S.,  Morgan
  E.~H.,   van~der Klis M.,  2002, \mn@doi [Monthly Notices of the Royal
  Astronomical Society] {10.1046/j.1365-8711.2002.05223.x}, 331, 745

\bibitem[\protect\citeauthoryear{Kuulkers, Lutovinov, Parmar, Capitanio,
  Mowlavi  \& Hermsen}{Kuulkers et~al.}{2003}]{Kuulkers2003}
Kuulkers E.,  Lutovinov A.,  Parmar A.,  Capitanio F.,  Mowlavi N.,   Hermsen
  W.,  2003, The Astronomer's Telegram, 149, 1

\bibitem[\protect\citeauthoryear{L{\"{a}}ngkvist, Karlsson  \&
  Loutfi}{L{\"{a}}ngkvist et~al.}{2014}]{Langkvist2014}
L{\"{a}}ngkvist M.,  Karlsson L.,   Loutfi A.,  2014, \mn@doi [Pattern
  Recognition Letters] {10.1016/j.patrec.2014.01.008}, 42, 11

\bibitem[\protect\citeauthoryear{Mackenzie, Pichara  \& Protopapas}{Mackenzie
  et~al.}{2016}]{Mackenzie2016}
Mackenzie C.,  Pichara K.,   Protopapas P.,  2016, \mn@doi [The Astrophysical
  Journal] {10.3847/0004-637x/820/2/138}, 820, 138

\bibitem[\protect\citeauthoryear{Mahabal, Sheth, Gieseke, Pai, Djorgovski,
  Drake, Graham  \& Collaboration}{Mahabal et~al.}{2017}]{Mahabal2017}
Mahabal A.,  Sheth K.,  Gieseke F.,  Pai A.,  Djorgovski S.~G.,  Drake A.,
  Graham M.,   Collaboration t.~C.,  2017, in IEEE Symposium Series on
  Computational Intelligence (SSCI). pp 2757--2764,
  \mn@doi{10.1109/SSCI.2017.8280984}

\bibitem[\protect\citeauthoryear{Mart{\'{i}}n-Fern{\'{a}}ndez,
  Barcel{\'{o}}-Vidal  \& Pawlowsky-Glahn}{Mart{\'{i}}n-Fern{\'{a}}ndez
  et~al.}{2003}]{Martin-Fernandez2003}
Mart{\'{i}}n-Fern{\'{a}}ndez J.~A.,  Barcel{\'{o}}-Vidal C.,   Pawlowsky-Glahn
  V.,  2003, \mn@doi [Mathematical Geology] {10.1023/A:1023866030544}, 35, 253

\bibitem[\protect\citeauthoryear{Massaro, Capitanio, Feroci, Mineo, Ardito  \&
  Ricciardi}{Massaro et~al.}{2020}]{Massaro2020-II}
Massaro E.,  Capitanio F.,  Feroci M.,  Mineo T.,  Ardito A.,   Ricciardi P.,
  2020, \mn@doi [Monthly Notices of the Royal Astronomical Society]
  {10.1093/mnras/staa1125}, 496, 1697

\bibitem[\protect\citeauthoryear{Matsuoka et~al.,}{Matsuoka
  et~al.}{2009}]{Matsuoka2009}
Matsuoka M.,  et~al., 2009, Publications of the Astronomical Society of Japan,
  61, 999

\bibitem[\protect\citeauthoryear{McInnes, Healy  \& Melville}{McInnes
  et~al.}{2018}]{McInnes2018}
McInnes L.,  Healy J.,   Melville J.,  2018, {UMAP: Uniform Manifold
  Approximation and Projection for Dimension Reduction}, \url
  {http://arxiv.org/abs/1802.03426}

\bibitem[\protect\citeauthoryear{Mirabel \& Rodriguezt}{Mirabel \&
  Rodriguezt}{1994}]{Mirabel1994}
Mirabel I.~F.,  Rodriguezt L.~F.,  1994, \mn@doi [Nature]
  {https://doi.org/10.1038/371046a0}, 371, 46

\bibitem[\protect\citeauthoryear{Motta}{Motta}{2016}]{Motta2016}
Motta S.~E.,  2016, \mn@doi [Astronomische Nachrichten]
  {10.1002/asna.201612320}, 337, 398

\bibitem[\protect\citeauthoryear{Naik, Agrawal, Rao  \& Paul}{Naik
  et~al.}{2002}]{Naik2002}
Naik S.,  Agrawal P. P.~C.,  Rao P. A.~R.,   Paul B.,  2002, \mn@doi [Monthly
  Notices of the Royal Astronomical Society]
  {https://doi.org/10.1046/j.1365-8711.2002.05077.x}, 330, 487

\bibitem[\protect\citeauthoryear{Naul, Bloom, P{\'{e}}rez  \& Van
  Der~Walt}{Naul et~al.}{2018}]{Naul2018}
Naul B.,  Bloom J.~S.,  P{\'{e}}rez F.,   Van Der~Walt S.,  2018, \mn@doi
  [Nature Astronomy] {10.1038/s41550-017-0321-z}, 2, 151

\bibitem[\protect\citeauthoryear{Nayakshin, Rappaport  \& Melia}{Nayakshin
  et~al.}{2000}]{Nayakshin2000}
Nayakshin S.,  Rappaport S.,   Melia F.,  2000, \mn@doi [The Astrophysical
  Journal] {10.1086/308860}, 535, 798

\bibitem[\protect\citeauthoryear{Pahari \& Pal}{Pahari \&
  Pal}{2010}]{Pahari2010}
Pahari M.,  Pal S.,  2010, \mn@doi [Monthly Notices of the Royal Astronomical
  Society] {10.1111/j.1365-2966.2010.17378.x}, 409, 903

\bibitem[\protect\citeauthoryear{Pattnaik, Sharma, Alabarta, Altamirano,
  Chakraborty, Kembhavi, M{\'{e}}ndez  \& Orwat-Kapola}{Pattnaik
  et~al.}{2021}]{Pattnaik2021}
Pattnaik R.,  Sharma K.,  Alabarta K.,  Altamirano D.,  Chakraborty M.,
  Kembhavi A.,  M{\'{e}}ndez M.,   Orwat-Kapola J.~K.,  2021, \mn@doi [Monthly
  Notices of the Royal Astronomical Society] {10.1093/mnras/staa3899}, 501,
  3457

\bibitem[\protect\citeauthoryear{Pedregosa et~al.,}{Pedregosa
  et~al.}{2011}]{Pedragosa2011}
Pedregosa F.,  et~al., 2011, Journal of Machine Learning Research, 12, 2825

\bibitem[\protect\citeauthoryear{Pieringer, Pichara, Catel{\'{a}}n  \&
  Protopapas}{Pieringer et~al.}{2019}]{Pieringer2019}
Pieringer C.,  Pichara K.,  Catel{\'{a}}n M.,   Protopapas P.,  2019, \mn@doi
  [Monthly Notices of the Royal Astronomical Society] {10.1093/mnras/stz106},
  484, 3071

\bibitem[\protect\citeauthoryear{Pursiainen et~al.,}{Pursiainen
  et~al.}{2020}]{Pursiainen2020}
Pursiainen M.,  et~al., 2020, \mn@doi [Monthly Notices of the Royal
  Astronomical Society] {10.1093/mnras/staa995}, 494, 5576

\bibitem[\protect\citeauthoryear{Richards et~al.,}{Richards
  et~al.}{2011}]{Richards2011}
Richards J.~W.,  et~al., 2011, \mn@doi [The Astrophysical Journal]
  {10.1088/0004-637X/733/1/10}, 733

\bibitem[\protect\citeauthoryear{Rokach \& Maimon}{Rokach \&
  Maimon}{2005}]{Rokach2005}
Rokach L.,  Maimon O.,  2005, in , The Data Mining and Knowledge Discovery
  Handbook.
Springer, pp 321--352

\bibitem[\protect\citeauthoryear{Shallue \& Vanderburg}{Shallue \&
  Vanderburg}{2018}]{Shallue2018}
Shallue C.~J.,  Vanderburg A.,  2018, \mn@doi [The Astronomical Journal]
  {10.3847/1538-3881/aa9e09}, 155, 94

\bibitem[\protect\citeauthoryear{Singh \& Yassine}{Singh \&
  Yassine}{2018}]{Singh2018}
Singh S.,  Yassine A.,  2018, \mn@doi [Energies] {10.3390/en11020452}, 11

\bibitem[\protect\citeauthoryear{Valenzuela \& Pichara}{Valenzuela \&
  Pichara}{2018}]{Valenzuela2018}
Valenzuela L.,  Pichara K.,  2018, \mn@doi [Monthly Notices of the Royal
  Astronomical Society] {10.1093/mnras/stx2913}, 474, 3259

\bibitem[\protect\citeauthoryear{Virtanen et~al.,}{Virtanen
  et~al.}{2020}]{Virtanen2020}
Virtanen P.,  et~al., 2020, \mn@doi [Nature Methods]
  {10.1038/s41592-019-0686-2}, 17, 261

\bibitem[\protect\citeauthoryear{Wiseman et~al.,}{Wiseman
  et~al.}{2020}]{Wiseman2020}
Wiseman P.,  et~al., 2020, \mn@doi [Monthly Notices of the Royal Astronomical
  Society] {10.1093/mnras/staa2474}, 498, 2575

\bibitem[\protect\citeauthoryear{Yu, Si, Hu  \& Zhang}{Yu
  et~al.}{2019}]{Yu2019}
Yu Y.,  Si X.,  Hu C.,   Zhang J.,  2019, {A review of recurrent neural
  networks: Lstm cells and network architectures},
  \mn@doi{10.1162/neco{\_}a{\_}01199}

\bibitem[\protect\citeauthoryear{Zhang \& Bloom}{Zhang \&
  Bloom}{2021}]{ZhangBloom2021}
Zhang K.,  Bloom J.~S.,  2021, \mn@doi [Monthly Notices of the Royal
  Astronomical Society] {10.1093/mnras/stab1248}, 505, 515

\bibitem[\protect\citeauthoryear{Zhu, Yu, Wang, Ning  \& Tang}{Zhu
  et~al.}{2019}]{Zhu2019}
Zhu L.,  Yu F.~R.,  Wang Y.,  Ning B.,   Tang T.,  2019, {Big Data Analytics in
  Intelligent Transportation Systems: A Survey},
  \mn@doi{10.1109/TITS.2018.2815678}

\makeatother
\end{thebibliography}



\appendix

\section{Architecture of LSTM-VAE network}
\label{sec: network architecture}

\begin{figure}
	\includegraphics[width=\columnwidth]{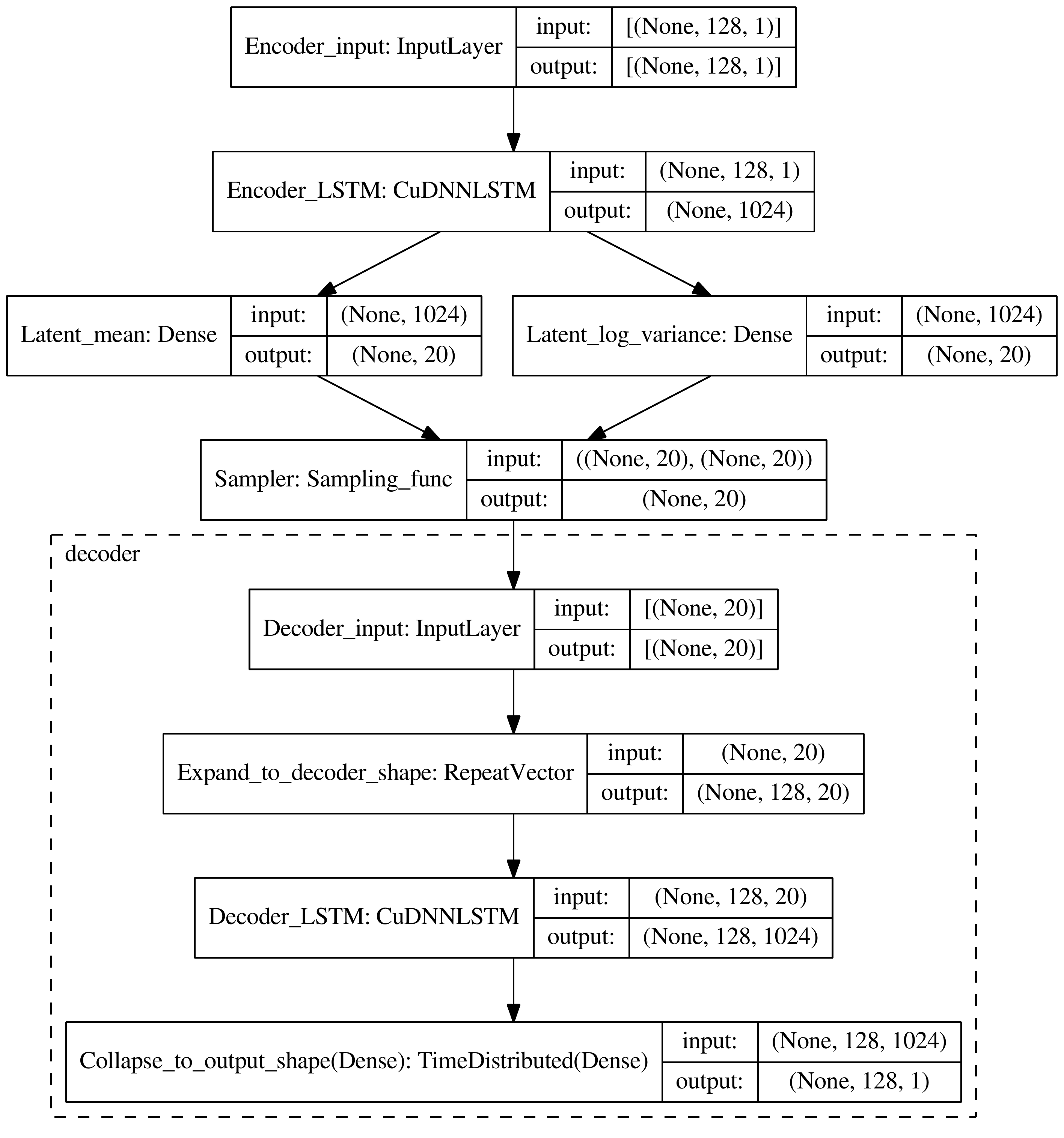}
    \caption{Architecture of the proposed LSTM-VAE model. Figure was generated using the Keras utility \(\texttt{plot\_model}\). Left-most cell of each block contains the label assigned to corresponding instance of a Keras object, followed by class of Keras model layer (except from \(\texttt{Sampling\_func}\) which is a custom function; see Appendix~\ref{sec: network architecture}}). Right-most cells contain shapes of input and output tensors of the objects. Shapes are presented using the convention followed by Keras.
    \label{fig:model plot}
\end{figure}

Figure~\ref{fig:model plot} shows a visualisation of the proposed LSTM-VAE architecture. The purpose of each layer is as listed below.

\begin{itemize}
    \item \(\texttt{Encoder\_input}\) creates an instance of a tensor with dimensions of the model input, i.e. a sequence of 128 values.
    \item \(\texttt{Encoder\_LSTM}\) is a layer of CuDNNLSTM cells. CuDNN stands for the CUDA Deep Neural Network library, which was developed by NVIDIA \citep{Chetlur2014}. The library accelerates training of neural networks using a graphical processing unit (GPU). This layer consists of 1024 such LSTM cells, which are not interconnected, but perform recurrent computation on the input sequence, one data point at a time. Output from every point of the sequence is stored within the state of the cell and used as input of the next computation of the sequence. Output of this layer consists of the final state of the cells, produced after the entire sequence has been processed. LSTM cells are trained to extract informative variables from the data through the process of backpropagation of errors. Increasing the number of LSTM cells tends to improve network's reconstruction loss (see below for the definition of reconstruction loss), and the number of 1024 is selected due to the GPU memory size constrain. 
    \item \(\texttt{Latent\_mean}\) and \(\texttt{Latent\_log\_variance}\) are two separate layers of fully interconnected neurons. Their purpose is to perform the dimensionality reduction of the 1024 variables extracted by the \(\texttt{Encoder\_LSTM}\) layer. \(\texttt{Latent\_mean}\) and \(\texttt{Latent\_log\_variance}\) each output 20 values (which is the dimensionality of the latent space). First set of 20 values is used as the mean of the continuous latent variables, whereas the other set encodes their spread. 
    In other words, \(\texttt{Encoder\_LSTM}\), \(\texttt{Latent\_mean}\) and \(\texttt{Latent\_log\_variance}\) make up the Encoder block of the VAE, which maps the network input to the mean and (log) variance vectors. Increasing the number of latent variables tends to improve network's reconstruction loss, but it also increases the dimensionality of the latent space and the computational complexity of downstream processes. We choose the number of latent variables based on the results of preliminary experiments, where increasing the number tended to improve the reconstruction quality, and the number of 20 reached the compromise between the reconstruction performance and the complexity of the latent space.
    \item \(\texttt{Sampler}\) generates random numbers from normal distributions whose parameters are set to the values of latent variable mean and variance. It is required in order to allow for deterministic treatment of the inherently probabilistic network during training (the so-called ``re-parameterisation trick'' \citep{Kingma2014}).
    \item \(\texttt{Decoder\_input}\) initialises the input tensor of the Decoder block of the model.
    \item \(\texttt{Expand\_to\_decoder\_shape}\) replicates the values of latent variables to create sequences of the same length as the initial light curve sequences. In other words, each LSTM cell of the Decoder block receives values of the 20 latent variables at each iteration of the sequential computation.
    \item \(\texttt{Decoder\_LSTM}\) layer is the Decoder counterpart of \(\texttt{Encoder\_LSTM}\) layer of the Encoder. It also performs recurrent computation on the sequential input, but rather than processing a single sequence of variable values, it processes 20 sequences of constant values. This layer has 1024 cells, each producing a sequence of cell states from each iteration of recurrent computation.
    \item \(\texttt{Collapse\_to\_output\_shape}\) is a fully connected layer which combines the 1024 sequences from the LSTM layer into a single sequence of 128 data points. Output of this layer is the reconstruction of an input light curve segment.
\end{itemize}

\section{Training and fine tuning of LSTM-VAE networks}
\label{sec: network training}

\begin{table}
	\centering
	\caption{Summary of the LSTM-VAE training and fine tuning. Validation loss stopped improving after the quoted number of training epochs. Training was stopped after 50 consecutive epochs without validation loss improvement. Best validation loss is shown.}
	\label{tab:training history}
	\begin{tabular}{lccc} 
		\hline
		Optimiser (rate) & \# epochs & Loss & Data set\\
		\hline
		Adam (Default) & 176 & 14.44 & 1s\\
		SGD (\(3\cdot10^{-4}\)) & 43 & 14.20 & \\
		SGD (\(1.5\cdot10^{-4}\)) & 112 & 14.17 & \\
		SGD (\(7.5\cdot10^{-5}\)) & 164 & 14.16 & \\
		\hline
		Adam (Default) & 209 & 107.63 & 4s\\
		SGD (\(3\cdot10^{-4}\)) & 5 & 103.67 & \\
		SGD (\(1.5\cdot10^{-4}\)) & 4 & 102.55 & \\
		SGD (\(7.5\cdot10^{-5}\)) & 12 & 102.02 & \\
		\hline
		
	\end{tabular}
\end{table}

\begin{figure}
	\includegraphics[width=\columnwidth]{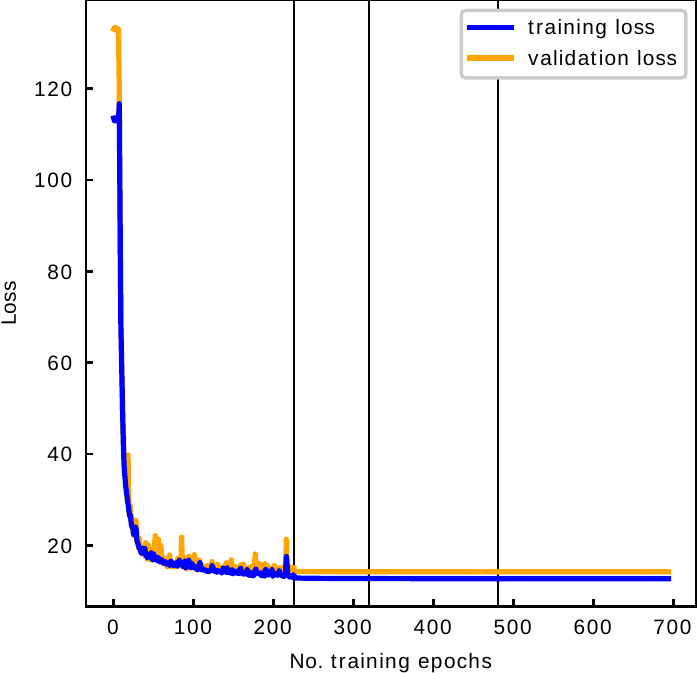}
	\includegraphics[width=\columnwidth]{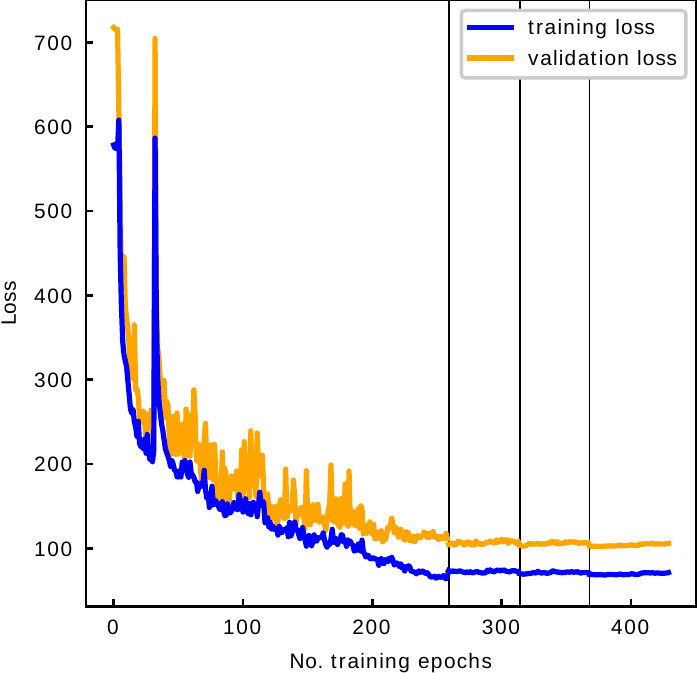}
    \caption{Loss curves resulting from the training of LSTM-VAE models on 1s data set (top) and on the 4s data set (bottom). Black vertical lines indicate that no improvement in validation loss in the last 50 consecutive epochs and change of the optimiser. See Table~\ref{tab:training history} for full training history.}
    \label{fig:loss curve}
\end{figure}

We train networks using the Keras implementation of Adam optimisation algorithm \citep{Kingma2015}, and fine-tune them using the stochastic gradient descend (SGD) algorithm. The \texttt{clipvalue} argument is set to 0.5 for both optimisers, which prevents numerical errors due to exploding gradients. Training is performed with the batch size of 1024 (number of light curve segments propagated through the network simultaneously) for various numbers of epochs (i.e. complete passes through the training set). Training is terminated when the validation loss value does not improve for 50 consecutive epochs. See Table~\ref{tab:training history} for full training history, and Figure~\ref{fig:loss curve} for loss curves resulting from the training. Results suggest that satisfactory training results could be achieved using Adam optimiser only, because improvements caused by SGD with decaying learning rate are marginal. \par

We use an Nvidia Geforce Titan Xp GPU for network training. One epoch of training takes $\sim$266 seconds of computation, resulting in the total of $\sim$36.6 and $\sim$17.0 hours of training for 1s and 4s data sets respectively, which could be reduced to $\sim$13.0 and $\sim$15.4 hours respectively, if fine tuning with SGD optimiser was not performed.\par

The amount of GPU memory (12 GB for Titan Xp) sets a limit on the number of parameters of the network for a given size of data batch. The number of parameters depends on the size of propagated tensors, which in turn depends on the sizes of layers of the model (see shapes of the Keras objects shown in Figure~\ref{fig:model plot}). In order to optimise the reconstruction performance within the constraints of the GPU memory limit, we maximise the number of LSTM cells, which increases the computational power of the network. Increasing the number of LSTM cells relative to the number of data points in the input layer (i.e. the segment length) tends to improve reconstruction performance. However, a decrease in the number of input data points means that less information about the context of variability patterns is given to the network, which shifts the focus of the analysis to short term patterns. The size of the training data batch could be reduced to allow for an increased number of parameters of the network, leading to improved performance at the expense of increased training time.

\section{LSTM-VAE reconstructions of 4s data set segments}
\label{sec:segment reconstructions 4s}
\begin{figure*}
	\includegraphics[width=\textwidth]{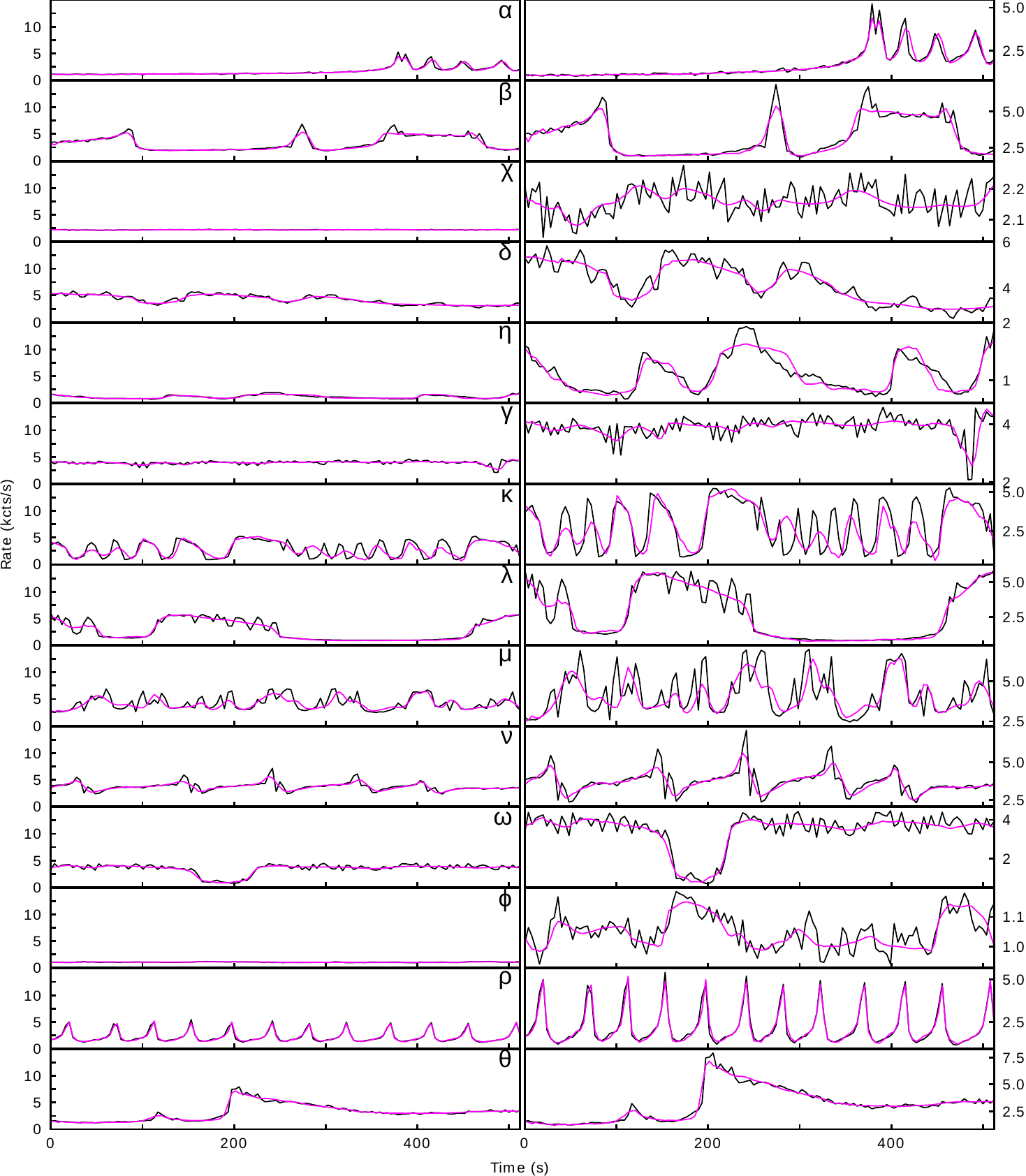}
    \caption{Examples of light curve segments and their LSTM-VAE reconstructions - the counterpart of Figure~\ref{fig:segment reconstructions 1s} for 4s data set.}
    \label{fig:segment reconstructions 4s}
\end{figure*}

See Figure~\ref{fig:segment reconstructions 4s} for example reconstructions of light curve segments from 4s data set.

\section{Gaussian mixture model}
\label{sec: gaussian mixture model}

GMM uses the Expectation-Maximisation \citep{Dempster1977} algorithm to approximate the probability distribution of the data using a set of multidimensional Gaussian components. The number of components is a hyper-parameter set by the user, and the mean position of each component is initiated randomly. Position and co-variance matrix of each component are then iteratively optimised to maximise the likelihood of the data under the model. GMM is a ``soft'' clustering method; likelihood value of each data point is calculated for each Gaussian component, and the data points are assigned to components which give the largest likelihood.\par

The grid search for the optimal number of components of the GMM is the most computationally expensive procedure described in this paper, barring the training of neural networks. The convergence time for this algorithm is highly dependent on the initial random state, but as an approximate point of reference, we inform that the fitting of the 222 component model to 1s data set and 279 component model to 4s data set require 1.96 and 2.91 CPU hours respectively, using a CPU with clock speed of 3.40 GHz. Additionally, the fitting of 400 component models to 1s and 4s data sets require 3.42 and 3.80 CPU hours respectively.\par

We use the default set of GMM hyper-parameters as implemented in the scikit-learn library \citep{Pedragosa2011}, with the exception of the number of components. We therefore fit full covariance matrices for each one of the Gaussian components. The computation time could be reduced at the expense of fit quality if different type of covariance matrix was chosen. Convergence threshold and the number of iterations of the Expectation-Maximisation algorithm are other tunable parameters.\par

\section{Random forest classifier}
\label{sec: random forest classifier}


The random forest classifier \citep{Breiman2001} uses an ensemble of decision trees to perform classification, and it averages the results across the trees to find the final classification. Each decision tree is trained on a sub-sample of observations, picked at random from the training data set using bootstrapping. Additionally, only a subset of features is used to train each decision tree, which is another source of randomness impacting the construction of the classifier. This randomness allows random forest to reduce over-fitting and the overall variance of the classifier, which leads to improved classification performance.\par

The scikit-learn \citep{Pedragosa2011} implementation of the random forest classifier controls the number of decision trees in the random forest ensemble using the n\_estimators hyper-parameter. Increasing the number of trees tends to reduce the variance of predictions, as the forest converges on the answer, but at the expense of increased computation time. Therefore, we keep this number constant for all the grid-search classifiers, and set the hyper-parameter to 1000 for the final  classifier used on the testing data subset.\par

Hyper-parameters min\_samples\_split and min\_samples\_leaf control the construction process of the decision trees, and we keep the values constant at 2 and 1 respectively. These hyper-parameters control the splitting of the trees, and increasing the values would not allow for the classes with just a single observation to be separated from other classes. We set the class\_weight hyper-parameter to ``balanced'', which sets the weights of the classes to be inversely proportional to the class frequency in order to account for the imbalance in the number of observations of each class.\par


Hyper-parameter criterion is used to choose between the functions which control the quality of a split in a tree, and the possible values of criterion are “gini” and “entropy”, which stand for Gini impurity and information gain respectively. Gini Index impurity criterion is defined as

\begin{equation}
    \text{I}_{G} = 1 - \sum_{k=1}^{K} p_k^{2}
	\label{eq:gini index}
\end{equation}
where \(K\) is the number of classes and \(p_k\) is the proportion of observations of class \(k\) in a set. Information gain is the difference between the entropy of a set before and after a split. Entropy is defined as

\begin{equation}
    \text{E} = -\sum_{k=1}^{K} p_k log(p_k)
	\label{eq:entropy}
\end{equation}
where \(K\) is the number of classes and \(p_k\) is the proportion of observations of class \(k\) in a set.\par
Any hyper-parameters of the random forest classifier which are not mentioned in the text were set to default values.\par

\section{Classification test with all available data}
\label{sec: validation classifications}

\begin{figure}
	\includegraphics[width=\columnwidth]{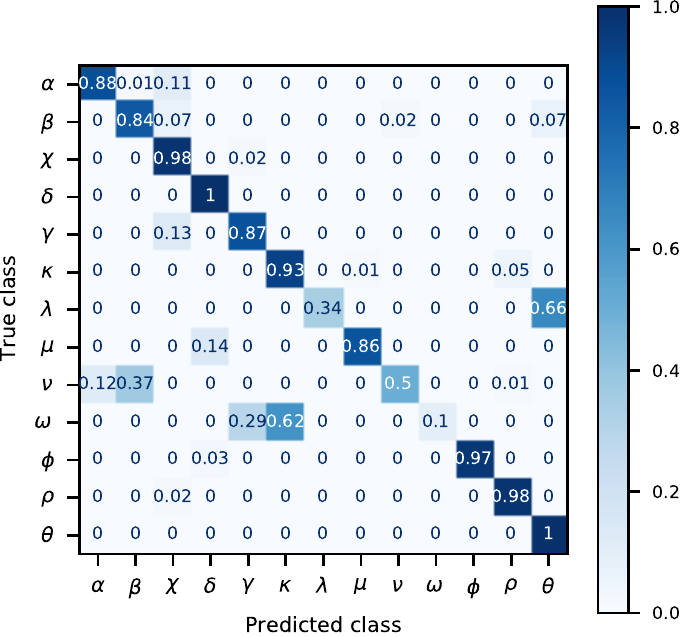}
    \caption{A row-wise normalised confusion matrix showing classification results of 100 test trials where 69 test observations are randomly sampled in a stratified manner from all of the available 206 labelled observations, as described in Section~\ref{sec: classification}. Rows of the matrix correspond to human-assigned classes (assumed true class), columns correspond to the classifications predicted by the classifier, and the values in the matrix represent the number of observations.  The numbers in each row are divided by the sum of numbers in that row.}
    \label{fig: train/val confusion matrix}
\end{figure}

Figure~\ref{fig: train/val confusion matrix} shows a summary of the classification results for the 100 tests where 69 test observations are randomly sampled in a stratified manner from the entire data set of 206 labelled observations (combined training, validation and test data subsets), and the remaining observations are used for training (as described in  Section~\ref{sec: classification}). The mean of weighted F1 and accuracy performance scores of those tests are \(0.877\pm0.031\) and \(0.896\pm0.026\) respectively. Observations of the majority of the classes are classified in agreement with the human-assigned classifications >\(80\%\) of the time, however observations of classes \(\lambda\), \(\nu\) and \(\omega\) are usually assigned to other classes. These classes are affected by very small sample size (each has only one training observation), which limits the statistical significance of their classification results.

\section{Hierarchical clustering}
\label{sec: hierarchical clustering}

Hierarchical (agglomerative) clustering algorithm creates a hierarchy of clustering solutions through recursive merging of pairs of clusters. The merging algorithm starts with the value of the distance threshold set to zero, resulting in the assignment of each data point to singleton clusters. The threshold is then increased, and pairs of nearby clusters are merged when the distance between them is less than the threshold. The algorithm finishes when all data points belong to a single cluster, or when a threshold set by the user is reached (see \citet{Rokach2005} for more details).\par

The hierarchical clustering algorithm can use a variety of distance metrics and different methods of calculating distance between clusters. We use the Euclidean metric and ``ward'' method, which uses the Ward variance minimisation algorithm and creates clusters of fairly regular sizes. The choice of clustering method affects the resulting cluster structure, for example ``single'' defines the distance between two clusters as the distance between the nearest data points of the two clusters, and it tends to build clusters with very uneven sizes, whereas ``complete'' method uses the distance between two furthest data points, and it tends to create more even, compact clusters.


\bsp	
\label{lastpage}
\end{document}